\documentstyle[aps,epsf]{revtex}
\begin{document}
\twocolumn[\hsize\textwidth\columnwidth\hsize\csname@twocolumnfalse\endcsname
\title{Nonlinear Elasticity of the Sliding Columnar Phase
}
\author{C. S. O'Hern\cite{email} and T. C. Lubensky}
\address{Department of Physics and Astronomy, University of Pennsylvania,
Philadelphia PA 19104}
\date{\today}
\maketitle
\begin{abstract}
The sliding columnar phase is a new liquid-crystalline phase of matter
composed of two-dimensional smectic lattices stacked one on top of the
other.  This phase is characterized by strong orientational but weak
positional correlations between lattices in neighboring layers and 
a vanishing shear modulus for sliding lattices relative to each other.  A
simplified elasticity theory of the phase only allows intralayer
fluctuations of the columns and has three important elastic constants:
the compression, rotation, and bending moduli, $B$, $K_y$, and $K$.
The rotationally invariant theory contains anharmonic terms
that lead to long wavelength renormalizations of the elastic 
constants similar to the Grinstein-Pelcovits renormalization of 
the elastic constants in smectic liquid crystals.  We calculate these
renormalizations at the critical dimension $d=3$ and find that 
$K_y(q) \sim K^{1/2}(q) \sim B^{-1/3}(q) \sim (\ln(1/q))^{1/4}$,
where $q$ is a wavenumber.  
The behavior of $B$, $K_y$, and $K$ in a model
that includes fluctuations perpendicular to the layers is identical to
that of the simple model with rigid layers.  We use dimensional
regularization rather than a hard-cutoff renormalization scheme
because ambiguities arise in the one-loop integrals with a finite
cutoff.
\\ {\sl Pacs: 61.30.Cz, 64.60.Ak,
87.10.+e.}
\end{abstract}
\vskip2pc]
\par

\section{Introduction}
\label{sec:introduction}

\begin{figure}
\epsfxsize=3.8truein
\centerline{\epsfbox{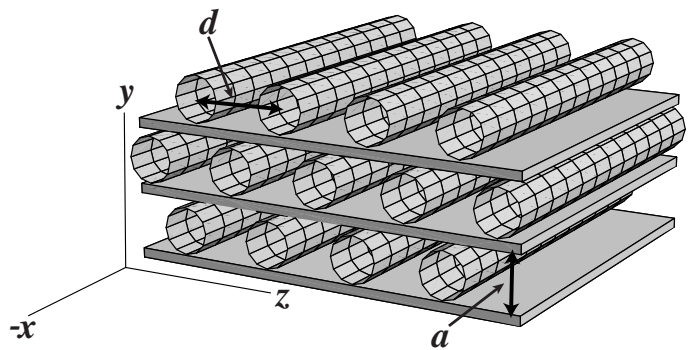}}
\caption{
Picture of the idealized sliding columnar phase.  The DNA columns
are sandwiched between planar lipid bilayer sheets.  The bilayer planes 
are stacked in the $y$-direction with spacing $a$.  The DNA columns are 
oriented in the $x$ direction, and, within each layer, the columns 
are separated by $d$.  The positions of columns in 
neighboring layers are uncorrelated.
}
\label{scfig}
\end{figure}

DNA, which is a semi-flexible polymer, and cationic lipids in solution
form complexes in which the negative charge of the DNA is nearly
compensated by the positive charge of the lipids.  These complexes are
under intensive study as possible nonviral carriers of DNA to cell
nuclei for gene therapy\cite{Felgner}.  R\"{a}dler, {\it et al.} have
shown that under appropriate conditions the complexes self-assemble
into multi-lamellar structures\cite{Salditt}.  The lipids form stacked
bilayer sheets with DNA molecules intercalated in the galleries
between the bilayers as shown in Fig.~\ref{scfig}.  Each gallery is thick
enough to accommodate only one DNA molecule and its hydration layer.
Within each gallery, DNA molecules adopt a linear rather than a coiled
configuration and form a regularly spaced parallel array that in the
absence of couplings to DNA in other galleries is a two-dimensional
smectic liquid crystal\cite{Wang}.  The experimentally determined
X-ray structure factor of these complexes is well modeled by a stack
of weakly coupled $2$D smectic lattices\cite{Salditt}.

Two recent theoretical papers\cite{Ohern,Golubovic} have pointed out
that weakly coupled $2$D smectic lattices form a new phase of matter,
the {\sl sliding columnar} phase.  This phase is characterized by strong
orientational correlations but weak positional correlations between
smectic lattices.  All lattices are aligned on average along a common
direction (the $x$-axis in Fig.~\ref{scfig}), but their relative
positions decorrelate exponentially with distance between smectic
lattices.  With sufficiently strong coupling between galleries,
long-range positional correlations between smectic layers develop, and
the system becomes an anisotropic columnar phase with a
two-dimensional DNA lattice in the plane perpendicular to the
direction of DNA alignment.  The sliding columnar phase, on the other
hand, is what the columnar phase becomes when coupling between
galleries becomes so weak that DNA lattices can slide freely across
each other.  It has no shear modulus resisting relative displacements
of DNA lattices, but it does have a rotation modulus resisting their
relative rotation.  Dislocations may melt the sliding columnar phase
to an anisotropic nematic lamellar phase at length scales longer 
than an in-plane dislocation unbinding length\cite{Toner}.
It is possible, however, to choose interlayer interactions so that
the sliding columnar phase is the stable equilibrium phase at all length
scales\cite{Oherndisc}.

This paper will investigate the nonlinear elasticity of the 
sliding columnar (SC) phase.  Its principal purpose is to show that 
the nonlinear strains lead to a Grinstein-Pelcovits renormalization
of the elastic constants \cite{Grinstein} and not, as one could imagine, to 
the destruction of the sliding columnar phase itself.  The lipid 
bilayers, which we take to be aligned on average parallel to the 
$xz$ plane as shown in Fig.~\ref{scfig}, fluctuate like 
bilayers in any lamellar phase.  To understand correlations and 
fluctuations of the DNA smectic lattices, it is convenient to consider
first a model in which the lipid bilayers are rigid planes with
no fluctuations in the $y$-direction.  In this case, displacements of the DNA
lattices, which are aligned on average along the $x$-direction, are
restricted to the $z$-direction.

The rotationally invariant Landau-Ginzburg-Wilson Hamiltonian in units
of $k_B T$ for this system is
\begin{equation}
{\cal H} = {1 \over 2} \int d^3 x \left[B u_{zz}^2 
+ K_y(\partial_x \partial_y u_z)^2 + K(\partial_x^2 u_z)^2\right],
\label{introenergy}
\end{equation}
where $B$, $K_y$, and $K$ are the compression, rotation, and 
bending moduli divided by $k_B T$ and 
\begin{equation}
u_{zz} = \partial_z u_z - {1 \over 2}\left[(\partial_x u_z)^2 +
(\partial_z u_z)^2\right] 
\label{introstrain}
\end{equation} 
is the nonlinear Eulerian strain appropriate for the sliding 
columnar phase.  Note that ${\cal H}$ is invariant under
\begin{equation}
u_z'({\bf x}) \rightarrow u_z({\bf x}) + f(y).
\end{equation}    
It is this fact that ensures that nonlinearities do not 
destroy the sliding columnar phase.

The rotationally invariant strain $u_{zz}$ introduces
anharmonic terms into the Hamiltonian that lead to a Grinstein-Pelcovits
renormalization of $B$, $K_y$, and $K$.  The renormalized moduli 
scale logarithmically with $q$ at long wavelengths:
\begin{equation}
K_y(q) \sim K^{1/2}(q) \sim B^{-1/3}(q)
\sim \left[\ln\left({\mu \over q}\right)\right]^{1/4},
\label{introrenorm}
\end{equation}
where $\mu$ is a large momentum cutoff.  A complete model for the 
sliding columnar phase allows both lipid bilayers and smectic 
lattices to fluctuate.  This model also exhibits Grinstein-Pelcovits
renormalization of the elastic constants.  Table \ref{exponents}
lists the exponents describing singularities in the elastic constants
for both the $3$D smectic and sliding columnar phases.

\begin{table}
\caption{Comparison of the logarithmic scaling exponents for the elastic
moduli of the 3D smectic and sliding columnar phases.  At long
wavelengths the elastic moduli for both phases scale as
$\ln^{\alpha}[1/q]$ with $\alpha$ given below.}
\begin{tabular}{cccc}
Phase & $B$ & $K$ & $K_y$\\
 \tableline
3D smectic & $-4/5$ & $2/5$ & --\\
sliding columnar & $-3/4$ & $1/2$ & $1/4$
\end{tabular}
\label{exponents}
\end{table}
The evaluation of the above renormalization presented some unexpected
difficulties.  The continuum Hamiltonian in (\ref{introenergy}) is
formally invariant under arbitrary global rotations.  However, the
introduction of a hard cutoff breaks this rotational invariance just
as the introduction of a similar cutoff breaks gauge invariance in
gauge Hamiltonians\cite{Lee}.  Nevertheless, hard cutoff RG procedures
can with care be applied successfully to Hamiltonians with gauge 
\cite{gauge} or
rotation symmetries\cite{rotation}.  Indeed, the original
Grinstein-Pelcovits calculation of the logarithmic renormalization of
the smectic-$A$ elastic constants used a
hard-cutoff\cite{Grinstein}.  When we applied the popular
momentum-shell hard cutoff RG procedure \cite{momentumshell}
to the nonlinearities in the
sliding columnar phase, we encountered ambiguities that we were unable
to resolve.  We found that the values of the one-loop diagrams
depended on whether the external momentum was added to the top or the
bottom part of the internal loop.  Similar difficulties are not
encountered in the Grinstein-Pelcovits calculation.  To eliminate
these ambiguities, we switched to the dimensional regularization
procedure which explicitly preserves rotational invariance because the
cutoffs are infinite\cite{Amit}.

The remainder of the paper will be organized as follows: we first
rederive the results of Grinstein and Pelcovits in Sec.\
\ref{sec:rganal3dsmectic} using dimensional regularization.  Then 
in Sec.\ \ref{sec:scrganal}, we calculate the renormalization of the
sliding columnar elastic constants of the simplified theory using the
same scheme.  In Sec.\ \ref{sec:layerfluct} we relax the constraint of
rigid membranes and show that the membrane fluctuations do not modify
the scaling behavior of the elastic moduli of the rigid theory.  In
Appendices
\ref{app:smecticloop} and \ref{app:slideloop}, we evaluate the one-loop 
diagrams for the 3D smectic and simplified sliding columnar theories.
In Appendix \ref{app:cutoff} we show that ambiguities arise when a
finite cutoff is implemented to calculate the loop diagrams of the
sliding columnar theory.  Finally, in Appendix \ref{app:nonlinstrain}
we derive the nonlinear strains required for the rotationally invariant
theory of the sliding columnar phase in the presence of 
fluctuating membranes.

\section{RG Analysis of the 3D Smectic}
\label{sec:rganal3dsmectic}

The rotationally invariant elasticity theory for a smectic liquid
crystal contains nonlinear terms that renormalize the elastic
constants of the harmonic theory for all dimensions below three.
Grinstein and Pelcovits calculated the corrections to the elastic
constants of a 3D smectic using an RG analysis with a finite
wavenumber cutoff\cite{Grinstein}.  They found that the corrections to
both the compression and bending moduli are logarithmic in the
wavenumber $q$ with the former scaling to zero and the latter scaling
to infinity at long wavelengths.  Application of a hard cutoff RG
procedure to the sliding columnar phase leads to ambiguities with no
obvious resolution.  (See Appendix \ref{app:cutoff}.)  We, therefore,
employ a dimensional regularization procedure that sends the cutoff
to infinity and thereby preserves rotational invariance.  In this
section we rederive the Grinstein-Pelcovits results for a 3D
smectic using dimensional regularization.  This establishes the
language needed to calculate the renormalization in the sliding
columnar phase.

\subsection{Rotationally Invariant Theory}
\label{sec:smecticrotenergy}

A smectic in $d$ dimensions is characterized by a mass-density wave
with period $P = 2\pi/q_0$ along one dimension and by fluid-like order
in the other $d-1$ dimensions.  The phase of the mass density wave at
point ${\bf x} = ({\bf x}_{\perp},z)$ is $q_0 u({\bf x})$.  The elastic
Landau-Ginzburg-Wilson Hamiltonian for a smectic in units of $k_B T$
is
\begin{equation}
{\cal H} = {1 \over 2} \int d^dx \left[ B_{{\rm sm}}u_{zz}^2 + K_{{\rm
sm}} ({\mbox{\boldmath{$\nabla$}}}_{\perp}^2u)^2 \right],
\label{smectic}
\end{equation}
where ${\mbox{\boldmath{$\nabla$}}}_{\perp}$ is the gradient operator
in the $d-1$ subspace spanned by ${\bf x}_{\perp}$ and $B_{{\rm sm}}$ and
$K_{{\rm sm}}$ are, respectively, the compression and bending moduli
divided by $k_B T$.  The nonlinear Eulerian strain $u_{zz} =
\partial_z u - (1/2)({\mbox{\boldmath{$\nabla$}}}u)^2$ is invariant
with respect to uniform, rigid rotations of the smectic layers.  Below
we will drop the $(\partial_z u)^2$ term in $u_{zz}$ since its
inclusion leads to nonlinear terms that are irrelevant in the RG sense
with respect to the two quadratic terms in (\ref{smectic}).
Therefore, we will take
\begin{equation}
u_{zz} \approx \partial_z u - {1 \over 2}
({\mbox{\boldmath{$\nabla$}}}_{\perp} u)^2.
\label{smstrain}
\end{equation}

\subsection{Engineering Dimensions}
\label{sec:smecticdimanal}

To implement our RG procedure it is convenient to rescale parameters
so that $B_{\rm sm}$ is replaced by unity and the nonlinear form of
$u_{zz}$ is preserved.  To this end, we scale $u$ and ${\bf x}$ 
as follows:
\begin{eqnarray}
u = L_u \widetilde{u},~~z = L_z \widetilde{z},~{\rm and}~
{\bf x}_{\perp} = \widetilde{{\bf x}}_{\perp}.
\label{scaledef}
\end{eqnarray}
Note that ${\bf x}_{\perp}$ does not rescale.  Under these rescalings
we obtain
\begin{equation}
u_{zz} = L_u L_z^{-1}\left(\partial_{\tilde{z}}\widetilde{u} - {1 \over 2}
L_u L_z({\mbox{\boldmath{$\nabla$}}}_{\widetilde{\perp}} \widetilde{u})^2 \right).
\label{nonlinearscale}
\end{equation}
We require $u_{zz} = A \widetilde{u}_{zz}$ where $\widetilde{u}_{zz} =
\partial_{\tilde{z}}\widetilde{u} - (1/2)
({\mbox{\boldmath{$\nabla$}}}_{\widetilde{\perp}} \widetilde{u})^2$ is the rescaled 
nonlinear strain with the same form as (\ref{smstrain}).  This yields
$L_u = L_z^{-1}$ and $A = L_u^2$.  The coefficient of $\widetilde{u}_{zz}^2$
in the rescaled Hamiltonian is set to one with the choice
\begin{equation}
L_u = B_{\rm sm}^{-1/3}.
\label{Lu}
\end{equation}
The rescaled theory then becomes
\begin{equation}
\widetilde{{\cal H}} = {1 \over 2} \int d^d \widetilde{x} 
\left[ \widetilde{u}_{zz}^2 + {1 \over w} 
({\mbox{\boldmath{$\nabla$}}}^2_{\widetilde{\perp}} \widetilde{u})^2 \right]
\label{scaled}
\end{equation}
with 
\begin{equation}
w = {B_{\rm sm}^{1/3} \over K_{\rm sm}}.
\label{w}
\end{equation}  
For the remainder of Sec. \ref{sec:rganal3dsmectic} we will use the
free energy in (\ref{scaled}) but drop the tilde on the scaled
variables.  

We determine the dimensions of the scaled variables using
the engineering dimensions of $B_{\rm sm}$ and $K_{\rm sm}$.  The
dimension $d_A$ determines how $A$ scales with length $L$: $[A] =
L^{d_A}$.  From the respective dimensions $d_{B_{\rm sm}} = -d$ and
$d_{K_{\rm sm}} = 2 - d$ of $B_{\rm sm}$ and $K_{\rm
sm}$, we obtain $[L_u] = [L_z^{-1}] = L^{d/3}$.
Using these we find the following for the dimensions of the scaled
variables and the parameter $w$:
\begin{eqnarray}
\label{dimension}
\left[ u \right]  & = & \left[{L \over L_u}\right] = L^{\epsilon/3},~~
\left[ z \right] =  \left[{L \over L_z}\right] = L^{1 + d/3}, \\
\left[ x_{\perp} \right] & = & \left[{L \over L_{x_{\perp}}}\right] 
= L,~{\rm and}~\left[ w \right] = \left[{L^{-d/3} \over L^{2 - d}}\right] 
= L^{-2 \epsilon/3}, \nonumber
\end{eqnarray}
where $\epsilon = 3-d$.  Using these definitions one can easily verify
that both terms in (\ref{scaled}) are dimensionless.  $[w]$ scales as
$\mu^{2\epsilon/3}$ where $[\mu] = L^{-1}$, and it is, therefore, 
a relevant variable below $d=3$.  We introduce a
dimensionless coupling constant $g_0$ via
\begin{equation}
w = (g_0 \mu)^{2\epsilon/3}
\label{wcoupling}
\end{equation}
to display explicitly the length dependence of $w$.  The dimensions of the
coefficients of the $(\partial_z u)^3$ and $(\partial_z u)^4$ terms we
omitted are, respectively, $2d/3$ and $4d/3$.  These terms are
irrelevant and will be ignored in what follows.

The engineering dimensions in (\ref{dimension}) imply that there is 
an invariance of ${\cal H}$ under the transformation $\mu \rightarrow 
\mu b$ and 
\begin{equation}
u({\bf x}_{\perp}, z) = b^{d_u} u'({\bf x}_{\perp}', z'),
\label{sminvariance}
\end{equation}
where ${\bf x}_{\perp}' = b^{-1} {\bf x}_{\perp}$ and 
$z' = b^{-(1+d/3)} z$, {\it i.e.}
\begin{equation}
{\cal H}[u,w,\mu] = {\cal H}[u', wb^{2\epsilon/3},\mu].
\label{hinvariance}
\end{equation}
This in turn implies a scaling form for the position correlation function
$G({\bf x}_{\perp}, z) = \langle u({\bf x}_\perp,z)u(0,0) \rangle$
and its Fourier transform $G({\bf q})$.
We find
\begin{equation}
G({\bf x}_{\perp},z,w) = b^{2(1-d/3)}G({\bf x}_{\perp}',z',wb^{2\epsilon/3}),
\label{smg}
\end{equation}
and from this we obtain the vertex function $\Gamma({\bf q})
= G^{-1}({\bf q})$,
\begin{equation}
\Gamma({\bf q}_{\perp}, q_z,w) = b^{-2(1 + d/3)} 
\Gamma(b {\bf q}_{\perp}, b^{1 + d/3} q_z, wb^{2\epsilon/3}).
\label{smvertex}
\end{equation}
When $d=3$ this reduces to the scaling form
\begin{equation}
\Gamma({\bf q}_{\perp},q_z, w) = q^4_{\perp}
\Gamma\left(1, {q_z \over q^2_{\perp}}, w\right),
\label{threepropagator}
\end{equation}
which the harmonic vertex function $\Gamma = q_z^2 + w^{-1}q_{\perp}^4$
satisfies.

\subsection{RG Procedure}
\label{sec:smecticrgprocedure}

To calculate renormalized quantities, we seek a multiplicative procedure
that yields a renormalized Hamiltonian with the same form
as the original Hamiltonian, {\it i.e.}, a Hamiltonian that is a 
function of a renormalized nonlinear strain with the same form as
(\ref{smstrain}).  To preserve the form of the strains, it is 
necessary to rescale fields and lengths simultaneously.  The rescaling
that produced (\ref{scaled}) shows that the form of $u_{zz}$ is 
preserved if the rescaling coefficients of $u$ and $z$ are 
inverses of each other.  We, therefore, introduce a renormalization
constant ${\cal Z}$ and a renormalized displacement $u'$ such
that 
\begin{equation}
u({\bf x}) = {\cal Z}^{1/3} u'({\bf x}') = 
{\cal Z}^{1/3} u'({\bf x}_{\perp}, {\cal Z}^{1/3} z).
\label{renormscale}
\end{equation}
This implies that $u_{zz}({\bf x}) = {\cal Z}^{2/3}u_{zz}'({\bf
x}')$.  We also introduce a unitless renormalized coupling constant
$g$ and renormalization constant ${\cal Z}_g$ via
\begin{equation}
w^{3/2} = g\mu^{\epsilon}{\cal Z}_g {\cal Z}^{1/2},
\label{definew}
\end{equation}
where $\mu$ is an arbitrary wavenumber scale.  The renormalized
Hamiltonian then becomes 
\begin{equation}
{\cal H}' = {1 \over 2} \int d^d x'~\left[ {\cal Z}(u'_{zz})^2 + \left(
g \mu^{\epsilon} {\cal Z}_g\right)^{-2/3} 
	\left( {\mbox{\boldmath{$\nabla$}}}^2_{\perp '} u'\right)^2 \right].
\label{scaledh}
\end{equation}
We now follow standard procedures to evaluate ${\cal Z}(g)$ 
and ${\cal Z}_g(g)$\cite{Amit}.  The renormalized Hamiltonian in 
(\ref{scaledh}) determines the vertex function 
\begin{eqnarray}
\label{oneloop}
\Gamma({\bf q}) & = & q_z^2 + 
\left(g \mu^{\epsilon}\right)^{-2/3}q_{\perp}^4
+ ({\cal Z} - 1)q_z^2 \\
& & + (g\mu^{\epsilon})^{-2/3}\left({{\cal Z}_g}^{-2/3}
-1\right)q_{\perp}^4 + \Sigma({\bf q}) \nonumber
\end{eqnarray}
to one-loop order,  where $\Sigma({\bf q})$ is the one-loop 
diagrammatic contribution to $\Gamma({\bf q})$.  
We next impose the following conditions on the vertex function to enforce
the correct scaling behavior:
\begin{mathletters}
\begin{eqnarray}
\label{condition1}
\left.{d \Gamma \over dq_z^2}\right|_{q_z = \mu^2, q_{\perp}=0} & = & 1 \\
\label{condition2}
\left.{d \Gamma \over dq_{\perp}^4}\right|_{q_z = \mu^2,q_{\perp}=0}
& = & \left(g \mu^{\epsilon}\right)^{-2/3}.
\end{eqnarray}
\end{mathletters}
In Appendix \ref{app:smecticloop} we show that the diagrammatic 
contributions are the following: 
\begin{mathletters}
\begin{eqnarray}
\left.{d\Sigma({\bf q}) \over dq_z^2}\right|_{q_z=\mu^2,q_{\perp}=0} & =
& -{g \over 16 \pi \epsilon} \\
\left.{d\Sigma({\bf q}) \over dq_{\perp}^4}\right|_{q_z=\mu^2,q_{\perp}=0} & =
& (g \mu^{\epsilon})^{-2/3} {g \over 32 \pi \epsilon}.
\label{epsiloncorrect}
\end{eqnarray}
\end{mathletters}
Using the conditions on the vertex function we obtain the relations 
for the renormalization constants in terms of the one-loop diagrammatic
corrections.  The following relations are correct to lowest order
in $\epsilon$:
\begin{mathletters}
\begin{eqnarray}
{\cal Z} & = & 1 + {g \over 16 \pi \epsilon} \\
\label{Zs}
{\cal Z}_g & = & 1 + {3 g \over 64 \pi \epsilon}.
\label{Zgs}
\end{eqnarray}
\end{mathletters}

\subsubsection{Callan-Symanzik Equation}
\label{sec:smecticCS}

The renormalized vertex function $\Gamma_r({\bf q})$ satisfies a
Callan-Symanzik (CS) equation under a change of length scale $\mu$.
We obtain the renormalized elastic moduli from the solution to
this equation.  The original theory in (\ref{scaled}) did not depend
on the length scale $\mu$.  We can therefore write the bare vertex function
$\Gamma$ in terms of the renormalized vertex function $\Gamma_r$ and
find the differential equation obeyed by $\Gamma_r$.  Since the
variables $u$ and $z$ scale as $u'({\bf x}) = {\cal Z}^{1/3}u({\bf
x}')$ and $z' = {\cal Z}^{1/3} z$, the vertex function must scale as
\begin{equation}
\Gamma({\bf q}_{\perp}, q_z, w) = {\cal Z}^{-1/3} \Gamma_r({\bf q}_{\perp},
{\cal Z}^{-1/3} q_z, g, \mu).
\label{renormgamma}
\end{equation}
The CS equation is determined by the condition $\mu d\Gamma/d\mu = 0$.
Since the renormalized vertex function can have explicit 
as well as implicit $\mu$ dependence through the 
functions ${\cal Z}$ and $g$, the CS equation for $\Gamma_r$ has three terms:
\begin{equation}
\left[ \mu {\partial \over \partial \mu} - {\eta(g) \over 3}
\left(1 + q_z {\partial \over \partial q_z} \right) + \beta(g) {\partial
\over \partial g} \right]\Gamma_r = 0,
\label{CSsmectic}
\end{equation}
where 
\begin{mathletters}
\begin{eqnarray}
\beta(g) & = & \mu {dg \over d\mu}, \\
\eta(g) & = & \beta(g) {d\left(\ln{\cal Z}\right) \over dg},
\label{csdef}
\end{eqnarray}
\end{mathletters}
and $q_z \partial/\partial q_z = q_z'\partial/\partial q_z'$ with 
$q_z' = {\cal Z}^{-1/3} q_z$.  This equation can
be integrated to yield an equation for $\Gamma_r$ as a function of the 
length scale $\mu$.
\begin{eqnarray}
\label{solution}
\Gamma_r({\bf q}_{\perp},q_z, g, \mu) 
& = & \exp\left[{1 \over 3} \int_0^l 
\eta dl'\right] \\
& & \times \Gamma_r\left({\bf q}_{\perp}, \exp\left[{1 \over 3} \int_0^l 
\eta dl'\right] q_z, g(l), \mu_0\right), \nonumber
\end{eqnarray}
where $\mu/\mu_0 = e^l$, $\mu d/d\mu = d/dl$, and 
\begin{equation}
\beta(g) = - {d g(l) \over dl}.
\label{betatog}
\end{equation}  
At $l=0$ we have set $\Gamma_r(l=0) = \Gamma_r({\bf
q}_{\perp},q_z,g_0,\mu_0)$.  Now we must solve for $\beta$ and $\eta$
in terms of $g$ in order to obtain the renormalized vertex function.
To find $\beta(g)$, we note that 
\begin{equation}
{dw^{3/2} \over dl} = {d \over dl}\left(g \mu_0^{\epsilon}
e^{\epsilon l} {\cal Z}_g {\cal Z}^{1/2} \right) = 0.
\label{windependence}
\end{equation} 
From this relation we find $\beta(g) = - \epsilon/(d(\ln Q)/dg)$ where
$Q = g {\cal Z}_g {\cal Z}^{1/2}$.  We can then plug in the
relations for ${\cal Z}$ and ${\cal Z}_g$, and we determine
$\beta$ and $\eta$ to be the following:
\begin{mathletters}
\begin{eqnarray}
\beta(g) = {5 \over 64 \pi} g^2 -\epsilon g\\
\eta(g) = -{1 \over 16 \pi} g.
\label{betaeta}
\end{eqnarray}
\end{mathletters}
In three dimensions $\epsilon=0$.  In this case, integration of 
$dg/dl$ yields 
\begin{equation}
g(l) = {g_0 \over 1 + 5 g_0 l/(64 \pi)},
\label{g}
\end{equation}
where $g_0 \equiv g(0) = w^{3/2}$.  The remaining task is simple; we must
evaluate the arguments of the exponentials in (\ref{solution}) to
obtain the $l$ dependence of $\Gamma_r$.  Since $g \sim 1/l$, the
integral of $\eta$ will scale as $\ln l$ and the exponentials of the
integral of $\eta$ will give power-law dependence on $l$. We find that
\begin{eqnarray}
\label{definition}
\exp\left[ {1 \over 3} \int_0^l \eta(l') dl' \right] & = & 
\left[1 + {5 g_0 \over
64 \pi} l \right]^{-4/15} \\
& \equiv & \left[{g/g_0}\right]^{4/15}. \nonumber
\end{eqnarray}

\subsubsection{Renormalized Elastic Constants}
\label{sec:smecticelastic}

The scaling relations in (\ref{smvertex}) and (\ref{solution}) 
imply that $\Gamma_r$ satisfies
\begin{eqnarray}
\label{dcorrect}
\Gamma_r({\bf q}_{\perp}, q_z, g, \mu) & = & 
b^{-4}\left[g/g_0\right]^{4/15} \\
& & \times \Gamma_r(b 
{\bf q}_{\perp}, b^2 \left[g/g_0\right]^{4/15} q_z, g, \mu_0 b). \nonumber
\end{eqnarray}
We now choose the reference length scale $b = \mu_0^{-1} = (q_z^2 +
w^{-1} q_{\perp}^4)^{-1/4}
\equiv \left[h({\bf q})\right]^{-1}$.
This implies that
\begin{equation}
l = \ln \left[ {\mu \over h({\bf q}) }
\right]
\label{ldefine}
\end{equation}
since $\mu/\mu_0 = e^l$.
We find the scaling form of the renormalized
vertex function, 
\begin{eqnarray}
\Gamma_r & = & \left[h({\bf q})\right]^4 
\left[g/g_0\right]^{4/15} \Gamma_r\left( {q_{\perp} \over 
h({\bf q}) }, { q_z 
\left[g/g_0\right]^{4/15} \over
\left[h({\bf q})\right]^2}, g, 1\right) \nonumber \\
& = & g^{-{2/3}} \left[g/g_0\right]^{4/15} 
q_{\perp}^4
+ \left[g/g_0\right]^{4/5} q_z^2,
\label{answer}
\end{eqnarray}  
by squaring the term in the second slot of the renormalized 
vertex function and 
adding it to $g^{-2/3}$ times the fourth power of the term in the first slot.
We then plug in (\ref{g}) for $g$ and transform back to variables
with dimension to find the following 
expression for the renormalized vertex
function:
\begin{eqnarray}
\label{finalanswer}
\Gamma_r({\bf q}) & = & B_{\rm sm}\left(1 + {5 g_0 \over 64 \pi}
\ln\left[\overline{\mu} \over \overline{h}({\bf q}) \right] 
\right)^{-4/5} q_z^2 \\
& + & K_{\rm sm}\left(1 + {5 g_0 \over 64 \pi}
\ln\left[\overline{\mu} \over \overline{h}({\bf q}) \right] \right)^{2/5}
q_{\perp}^4, \nonumber
\end{eqnarray}
where $g_0 = B^{1/2}_{\rm sm} K^{-3/2}_{\rm sm}$, $\overline{\mu} =
\mu/B_{\rm sm}^{1/6}$, and $\overline{h}({\bf q}) = (q_z^2 + \lambda^2
q_{\perp}^4)^{1/4}$ with $\lambda^2 = K_{\rm sm}/B_{\rm sm}$.
$\overline{\mu}^2$ is a wavenumber $\Lambda \sim 1/a$ associated with the
short distance scale $a$.  We identify the renormalized
compression and bending moduli $B_{\rm sm}({\bf q})$ and $K_{\rm
sm}({\bf q})$ as the coefficients of the $q_z^2$ and $q_{\perp}^4$
terms respectively.  The renormalized elastic constants
scale as powers of logarithms at long wavelengths:
\begin{equation}
K_{\rm sm}({\bf q}) \sim B^{-1/2}_{\rm
sm}({\bf q})
\sim \left[\ln\left(\overline{\mu} 
\over \overline{h}({\bf q}) \right)\right]^{2/5},
\label{smexponents}
\end{equation} 
where the long wavelength regime is defined by wavenumbers $q$ that 
satisfy  $\overline{h}({\bf q}) \ll \Lambda^{1/2} 
\exp\left[-64\pi/(5g_0)\right]$.
We see that $K_{\rm sm}({\bf q})$ scales to
infinity and $B_{\rm sm}({\bf q})$ scales to zero as $q \rightarrow
0$.

\section{The Sliding Columnar Phase with Rigid Layers}
\label{sec:scrganal}

In this section we calculate the logarithmic corrections to the
elastic constants for the sliding columnar phase using the dimensional
regularization scheme employed in the previous section.  The steps we
follow for the dimensional regularization of the SC phase closely
resemble those followed for the dimensional regularization of the 3D
smectic phase since the two Hamiltonians have similar forms.  In this
section we assume that each 2D lattice of columns is flat and only
allowed to fluctuate in the $z$-direction.  We relax this assumption
in Sec. {\ref{sec:layerfluct} and find that the renormalized
elastic constants are identical to those of the flat theory to lowest
order in the coupling between strains in the $y$- and $z$-directions.

\subsection{Rotationally Invariant Theory}
\label{sec:scrotenergy}

The rotationally invariant elasticity theory describing the
sliding columnar phase was derived previously in
\cite{Ohern,Golubovic}.  We found that a
phase with weak positional correlations but strong orientational
correlations between neighboring $2$D smectic lattices was possible for
sufficiently low temperatures.  The strong orientational correlations
require a rotation modulus in the Landau-Ginzburg-Wilson Hamiltonian
that assesses an energy cost for relative rotations of the lattices in
addition to the compression and bending energy costs for a single
lattice of columns.  The Hamiltonian for the idealized sliding columnar
phase in three dimensions and in units of $k_B T$ is
\begin{equation}
{\cal H} = {1 \over 2} \int d^3x \left[ Bu_{zz}^2 + K(\partial_x^2
u_z)^2 + K_y (\partial_y \partial_x u_z)^2 \right],
\label{scfreeenergy}
\end{equation}
where $B$, $K_y$, and $K$ are the compression, rotation, and bending
moduli divided by $k_B T$.  Symmetry permits additional terms in the
Hamiltonian proportional to $K_{zy}(\partial_z \partial_y u_z)^2$ and
$K_{zx}(\partial_z \partial_x u_z)^2$.  The $K_{zy}$ term measures the
energy cost associated with variation in the DNA lattice spacing from
layer to layer, and the $K_{zx}$ term measures the energy cost
associated with the variation in the orientation with strand number of
DNA strands within a layer.  These terms are, however, subdominant to
those kept in (\ref{scfreeenergy}), and the couplings $K_{zy}$ and
$K_{zx}$ are irrelevant.  We will ignore them in what follows.  The
nonlinear strain $u_{zz}$ is identical to the nonlinear strain for one
layer of columns $u_{zz} =
\partial_z u_z - (1/2)[(\partial_x u_z)^2 + (\partial_z u_z)^2]$.
Below we will drop the $(\partial_z u_z)^2$ term from the nonlinear
strain since it leads to terms in the nonlinear theory that are also
irrelevant with respect to the three harmonic terms in
(\ref{scfreeenergy}).  Therefore, we use the approximate 
expression,
\begin{equation}
u_{zz} \approx \partial_z u_z - {1 \over 2}\left(\partial_x u_z\right)^2.
\label{scstrain}
\end{equation}

We note that $u_{zz}$ and ${\cal H}$ do not possess a shear strain
term $(\partial_y u_z)^2$ because neighboring layers of columns can
slide relative to one another without energy cost.  The absence of the
shear energy cost is a unique feature of the
sliding columnar elasticity theory.  Because the Hamiltonian lacks
terms with $y$-derivatives alone, it is invariant with respect to
shifts in $u_z$ that are only a function of $y$.  Hence, ${\cal
H}[u_z'] = {\cal H}[u_z]$ with
\begin{equation}
u_z' = u_z + f(y).
\label{invariance}
\end{equation}
This invariance restates that there is no energy cost for sliding
neighboring layers of columns relative to one another by an arbitrary
amount.

\subsection{Engineering Dimensions}
\label{sec:scdimanal}

We simplify the sliding columnar theory in
(\ref{scfreeenergy}) by rescaling the lengths so that $B$ and $K_y$
are replaced by unity and the nonlinear form of $u_{zz}$ is preserved.
We accomplish this by scaling $u_z$, $y$, and $z$ but not $x$.  To
implement a dimensional regularization scheme it is necessary to 
let $x$ become a $d-2$
dimensional displacement in the space perpendicular to $y$ and $z$.
Rescaled variables are defined via
\begin{equation}
\begin{array}{ll}
u_z = L_u \widetilde{u}_z, & x = \widetilde{x},\\
y = L_y \widetilde{y},~{\rm and} & z = L_z \widetilde{z}.
\label{scscaling}
\end{array}
\end{equation}
We first set $L_u = L_z^{-1}$ to preserve the form of $u_{zz}$ 
under (\ref{scscaling}).  
We then set the coefficients of 
$\widetilde{u}_{zz}^2$ and 
$(\partial_{\tilde{y}} \partial_{\tilde{x}} \widetilde{u}_z)^2$ to 
unity by choosing 
\begin{eqnarray}
L_y = \left({K_y^3 \over B}\right)^{1/4}~{\rm and}~
L_z = (K_y B)^{1/4}.  
\label{lylz}
\end{eqnarray}
The rescaled Hamiltonian becomes
\begin{equation}
\widetilde{{\cal H}} = {1 \over 2} \int d^d \widetilde{x}
\left[ \widetilde{u}_{zz}^2 + \left(\partial_{\tilde{x}}
\partial_{\tilde{y}} \widetilde{u}_z\right)^2 +
w^{-1} (\partial_{\tilde{x}}^2 \widetilde{u}_z)^2 \right]
\label{scaledsc}
\end{equation}
with 
\begin{equation}
w = {B^{1/2} \over K K_y^{1/2} }
\label{wsc}
\end{equation}  
and $d = 3-\epsilon$.  In the rest of this section we use
(\ref{scaledsc}) and drop the tildes.

We determine the dimension of the scaled variables from the dimensions
of the elastic constants in (\ref{scfreeenergy}).  The dimensions 
$[B] =
L^{-d}$ and $[K_y] = [K] = L^{2 - d}$ dictate
\begin{equation}
\begin{array}{lll}
\left[u_z\right] = L^{(3 - d)/2}, &
\left[x\right] = L, &  \\
\left[y\right] = L^{(d - 1)/2}, &
\left[z\right] = L^{(d + 1)/2},~{\rm and} &
\left[w\right] = L^{d - 3}.
\label{scdimension}
\end{array}
\end{equation}
Note that $[w]$ scales as $\mu^{\epsilon}$ with $[\mu] = L^{-1}$ and
is relevant below $d=3$.  The length dependence of $w$ is extracted by
introducing a dimensionless coupling constant $g_0$ via $w =
g_0\mu^{\epsilon}$.

The engineering dimensions in (\ref{scdimension}) imply that the 
Hamiltonian is invariant under the transformations $\mu \rightarrow \mu b$
and
\begin{equation}
u_z({\bf x}) = b^{d_u}u_z'({\bf x}')
\label{sctransform}
\end{equation}
with $x' = b^{-1}x$, $y' = b^{-(d-1)/2}y$, and $z' = b^{-(d+1)/2}z$,
{\it i.e.} the Hamiltonian obeys
\begin{equation}
{\cal H}\left[u_z,w,\mu\right] = {\cal H}\left[u_z', wb^{\epsilon},\mu\right].
\label{invarianth}
\end{equation}
This implies that there is a scaling form for the position correlation 
function $G({\bf x}) = \langle u_z({\bf x})u_z(0)\rangle$ and the 
vertex function $\Gamma = G^{-1}$.  We find that $\Gamma({\bf q})$ 
obeys the following scaling relation:
\begin{eqnarray}
\label{vertexscaling}
\Gamma({\bf q},w) & = & \\
& & b^{-(d+1)} \Gamma\left(bq_x, b^{(d-1)/2} 
q_y, b^{(d+1)/2} q_z,wb^{\epsilon}\right). \nonumber
\end{eqnarray} 
When $d=3$ this reduces to 
\begin{equation}
\Gamma({\bf q},w) = q_x^4 \Gamma\left(1, q_y/q_x, 
q_z/q_x^2\right),
\label{check}
\end{equation}
which is satisfied by the SC harmonic vertex function 
$\Gamma = q_z^2 + q_x^2q_y^2 + w^{-1}q_x^4$.

\subsection{RG Procedure}
\label{sec:scrgprocedure}

We now follow closely the RG procedure in
Sec. \ref{sec:smecticrgprocedure}.  We rescale the lengths and fields,
ensure that the SC Hamiltonian has the same form as the unscaled
SC Hamiltonian, impose boundary conditions on the vertex function, and
determine the renormalization constants in terms of the one-loop
diagrammatic corrections.  The first step in the process is to rescale
lengths such that the renormalized SC Hamiltonian has the same form as
(\ref{scaledsc}).  To preserve the form of the nonlinear strain, the
$z$ and $u$ rescalings must be inverses of one another and the $y$
rescaling is arbitrary.  We, therefore, introduce two renormalization
constants ${\cal Z}$ and ${\cal Z}_y$ such that
\begin{equation}
u_z({\bf x}) = {\cal Z}^{1/3} u_z'({\bf x}') = {\cal Z}^{1/3} 
u_z'(x,{\cal Z}_y y, {\cal Z}^{1/3} z).
\label{scscalerules}
\end{equation}
This implies that $u_{zz}({\bf x}) = {\cal Z}^{2/3} u_{zz}'({\bf x}')$ and
$\partial_x \partial_y u_z({\bf x}) = {\cal Z}^{1/3} {\cal Z}_y
\partial_{x'} \partial_{y'} u_z'({\bf x}')$.  
We also define a unitless coupling constant $g$ and 
renormalization constant ${\cal Z}_g$ by setting  
\begin{equation}
w = g\mu^{\epsilon} {\cal Z}^{1/3} {\cal Z}_g {\cal Z}^{-1}_y. 
\label{zgsc}
\end{equation}
The renormalized Hamiltonian then becomes
\begin{eqnarray}
\label{hamiltoniansc}
{\cal H}' & = & {1 \over 2} 
\int d^dx'\Big[ {\cal Z} {\cal Z}_y^{-1} (u_{zz}')^2 
+ {\cal Z}^{1/3} {\cal Z}_y (\partial_{x'} \partial_{y'} u_z')^2 \nonumber \\
& + & (g\mu^{\epsilon} {\cal Z}_g)^{-1}(\partial_{x'}^2u_z')^2\Big]. 
\end{eqnarray} 

We again employ standard RG procedures to calculate ${\cal Z}$, ${\cal Z}_y$,
and ${\cal Z}_g$.  The renormalization constants are fixed once we
impose the following three conditions on the 
vertex function:
\begin{eqnarray}
\label{rgcondsc1}
\left.{d\Gamma \over dq_z^2}\right|_{q_z = \mu^2, q_{x,y} = 0} & = & 1 \\
\left.{d\Gamma \over 
d(q_x^2 q_y^2)}\right|_{q_z = \mu^2, q_{x,y} = 0} & = & 1 \nonumber \\
\left.{d\Gamma \over dq_x^4}\right|_{q_z = \mu^2, q_{x,y} = 0} 
& = & (g\mu^{\epsilon})^{-1}. \nonumber
\end{eqnarray}
(Note that we have dropped the primes on the rescaled Hamiltonian.)  The 
vertex function to one-loop order,
\begin{eqnarray}
\Gamma & = & q_z^2 + q_x^2q_y^2 + (g\mu^{\epsilon})^{-1} q_x^4 
+ \left({\cal Z}{\cal Z}_y^{-1} - 1\right)q_z^2\\
& + & \left({\cal Z}^{1/3}
{\cal Z}_y - 1\right)q_x^2 q_y^2 
+ (g\mu^{\epsilon})^{-1}\left({\cal Z}_g^{-1}
-1\right)q_x^4 + \Sigma({\bf q}), \nonumber 
\end{eqnarray}
is obtained from (\ref{hamiltoniansc}) by adding and subtracting
$q_z^2 + q_x^2q_y^2 + (g\mu^{\epsilon})^{-1}q_x^4$ and including the
one-loop diagrammatic contributions to the vertex function,
$\Sigma({\bf q})$. In Appendix \ref{app:slideloop} we calculate the
diagrammatic contributions,
\begin{mathletters}
\begin{eqnarray}
\left.{d\Sigma \over dq_z^2}\right|_{q_z = \mu^2, q_{x,y} = 0} & = & 
-{g \over 8\pi^2 \epsilon} \\
\left.{d\Sigma \over d(q_x^2q_y^2)}\right|_{q_z = \mu^2, q_{x,y} = 0} & = &
{g \over 24 \pi^2 \epsilon} \\
\left.{d\Sigma \over dq_x^4}\right|_{q_z = \mu^2, q_{x,y} = 0} & = &
(g \mu^{\epsilon})^{-1} {g \over 12 \pi^2 \epsilon},
\label{slidesig}
\end{eqnarray}
\end{mathletters}
to lowest order in $\epsilon$.  From these we determine the
renormalization constants to be
\begin{mathletters}
\begin{eqnarray}
{\cal Z} & = & 1 + {g \over 16 \pi^2 \epsilon} \\
{\cal Z}_y & = & 1 - {g \over 16 \pi^2 \epsilon} \\
{\cal Z}_g & = & 1 + {g \over 12 \pi^2 \epsilon}.
\label{zanswer}
\end{eqnarray}
\end{mathletters}

\subsubsection{Callan-Symanzik Equation}
\label{sec:scCS}

The Callan-Symanzik equation is obtained by requiring
that the original theory in (\ref{scaledsc}) be independent of the
length scale $\mu$.  To ensure this, we set $\mu d\Gamma/d\mu = 0$.
This can be converted into a differential equation in the renormalized 
vertex function $\Gamma_r$ using the following scaling relation:
\begin{equation} 
\Gamma({\bf q},w) = {\cal Z}^{-1/3} {\cal Z}_y \Gamma_r\left(
q_x,{\cal Z}^{-1}_yq_y,{\cal Z}^{-1/3}q_z,g,\mu\right).
\label{scalinggamma}
\end{equation}
From the scaling relation we determine that the CS equation has
the following four terms:
\begin{eqnarray}
\Big[ \mu {\partial \over \partial \mu} - {\eta(g) \over 3}
\left(1 + q_z {\partial \over \partial q_z}\right) \nonumber \\
\label{CSsc}
+ \eta_y(g)
\left(1 - q_y {\partial \over \partial q_y} \right) + 
\beta(g) {\partial
\over \partial g} \Big]\Gamma_r = 0,
\end{eqnarray}
where $\eta(g)$ and $\beta(g)$ were defined previously in
Sec. \ref{sec:smecticCS} and $\eta_y(g) = \beta(g) d(\ln {\cal
Z}_y)/dg$.  The solution to (\ref{CSsc}) is 
\begin{eqnarray}
\label{scsolution}
& & \Gamma_r\left({\bf q},g, \mu\right) = \exp\left[
\int_0^l \left( {\eta \over 3} - \eta_y\right)  dl' \right] \times \\
& & \Gamma_r\left(q_x, \exp\left[\int_0^l \eta_y dl'\right] q_y, 
\exp\left[{1 \over 3} \int_0^l \eta dl'\right] q_z, g, \mu_0\right), \nonumber
\end{eqnarray}
with $\Gamma_r(l=0) = \Gamma_r({\bf q},g_0,\mu_0)$ and $\mu/\mu_0 = e^l$.

The coupling
constant $w$ must be independent of the length scale $l$.  This condition
yields a differential equation for the dimensionless constant 
$g$ whose solution is 
\begin{equation}
g(l) = {g_0 \over 1 + g_0 l/(6 \pi^2)}.
\label{gsc}
\end{equation}
This equation in turn determines the $l$ dependence of $\eta$ and
$\eta_y$ since they are both proportional to $g$.  We find
\begin{equation}
\eta(g) = -\eta_y(g) = {g \over 16 \pi^2},
\label{etaetay}
\end{equation}  
and thus these scale as $1/l$ at long wavelengths.

\subsubsection{Renormalized Elastic Constants}
\label{sec:scelastic}

Using (\ref{gsc}) for $g(l)$ and the relations for $\eta(g)$ and
$\eta_y(g)$ in (\ref{etaetay}), we obtain the scaling form of the
renormalized vertex function:
\begin{eqnarray}
\label{scalingsolution}
& & \Gamma_r({\bf q},g,\mu) =  
b^{-4}\left[g/g_0\right]^{1/2} \times \\
& & \Gamma_r\left(bq_x, bq_y 
\left[g/g_0\right]^{-3/8}, b^2q_z \left[g/g_0\right]^{1/8},
g, \mu_0b\right). \nonumber
\end{eqnarray}
To set the length scale, we choose
\begin{equation}
b = \mu_0^{-1} = (q_z^2 + q_x^2q_y^2 + w^{-1}q_x^4)^{-1/4} \equiv
\left[h({\bf q})\right]^{-1}.
\label{choose}
\end{equation}
It follows that
\begin{equation}
l = \ln\left[{ \mu \over h({\bf q})} \right]
\label{sclengthscale}
\end{equation} 
since $\mu$ and $l$ are related via $\mu/\mu_0 = e^l$.  
We then substitute (\ref{gsc}) for $g$ and
transform back to variables with dimension to obtain the following
expression for $\Gamma_r({\bf q})$:
\begin{eqnarray}
\label{scanswer}
\Gamma_r({\bf q}) & = & B\left(1 + {g_0 \over 6 \pi^2}
\ln\left[\overline{\mu} \over \overline{h}({\bf q})\right] 
\right)^{-3/4} q_z^2 \\
& + & K_y\left(1 + {g_0 \over 6 \pi^2}
\ln\left[\overline{\mu} \over \overline{h}({\bf q})\right] 
\right)^{1/4} q_x^2 q_y^2 \nonumber \\
& + & K\left(1 + {g_0 \over 6 \pi^2}
\ln\left[\overline{\mu} \over \overline{h}({\bf q})\right] \right)^{1/2}
q_x^4, \nonumber
\end{eqnarray} 
where $g_0 = B^{1/2}/(K K^{1/2}_y)$, $\overline{\mu} = \mu/(K_yB)^{1/8}$, and
$\overline{h}({\bf q}) = (q_z^2 + \lambda_y^2 q_x^2q_y^2 + \lambda^2
q_x^4)^{1/4}$ with $\lambda_y^2 = K_y/B$ and $\lambda^2 = K/B$.  
$\overline{\mu}^2$ is an upper momentum cutoff $\Lambda \sim 1/a$ 
associated with the short distance scale $a$.  We
can now identify the $q$ dependent elastic constants and determine
their scaling as $q$ tends to zero.  At long wavelengths such
that $\overline{h}({\bf q}) \ll \Lambda^{1/2} \exp\left[-6\pi^2/g_0\right]$
the $\ln$ term dominates, and we find
\begin{equation}
K_y({\bf q}) \sim K^{1/2}({\bf q}) \sim
B^{-1/3}({\bf q})
\sim \left[\ln\left({\overline{\mu} \over \overline{h}({\bf q})}\right)\right]^{1/4}.
\label{qdependent}
\end{equation}
We see that $B({\bf q})$ scales to zero and $K({\bf q})$ and $K_y({\bf q})$ 
scale to infinity as $q \rightarrow 0$.  Also note in Table
\ref{exponents} that the exponents of the logarithmic power-laws of
$B({\bf q})$ and $K({\bf q})$ are different from those of $B_{\rm
sm}({\bf q})$ and $K_{\rm sm}({\bf q})$, but the signs of the
respective exponents are the same.

\section{Sliding Columnar Phase with Fluctuating Lipid Bilayers}
\label{sec:layerfluct}

In the preceeding section, we considered a model for lamellar
DNA-lipid complexes in which lipid bilayers were treated as rigid
planes and no displacements of DNA lattices in the $y$-direction were
allowed.  In physically realized complexes, lipid bilayers can undergo
shape fluctuations and DNA lattices can undergo $y$-displacements.  We
can parameterize the shape of the $n$th bilayer by a height function
$h_n(x,z)$, which in the continuum limit becomes $h({\bf
x})=h_{y/a}(x,z)$.  The $y$-displacement of the DNA lattices in the
continuum limit is $u_y({\bf x})$.  At long wavelengths the
displacements $h({\bf x})$ and $u_y({\bf x})$ are locked together.
The lock-in occurs because there is an energy cost for translating
each lattice of columns and the lipid bilayers by different constant
amounts in the $y$-direction.  (See Fig.\ref{scfig}.)  We can,
therefore, describe long wavelength elastic distortions and
fluctuations of the sliding columnar phase in terms of a
Landau-Ginzburg-Wilson elastic Hamiltonian expressed in terms of
displacements $u_z$ and $u_y$:
\begin{eqnarray}
\label{uyuz}
& & {\cal H}_b\left[u_y,u_z\right] =  {1 \over 2} \int d^3x~\Big[
B^zu_{zz}^2 + K^z_{xx}(\partial^2_xu_z)^2 \\
& + & K^z_{xy}(\partial_x\partial_y u_z)^2
+ B^y u^2_{yy} + K^y_{xx}(\partial^2_x u_y)^2 \nonumber \\
& + & K^y_{xz}(\partial_x\partial_z u_y)^2 + 
K^y_{zz}(\partial^2_zu_y)^2 + 2 B^{yz}u_{yy}u_{zz}\Big], \nonumber
\end{eqnarray}
where $u_{yy}$ and $u_{zz}$ are nonlinear strains.  We define
${\cal H}_b$ to have units of $k_BT$, and therefore the constants
appearing in this equation are the compression and bending moduli
divided by $k_BT$.  The first three terms in (\ref{uyuz}) were discussed
previously in Sec. \ref{sec:scrganal} as the $u_z$ theory for the
sliding columnar phase without fluctuations of the lipid bilayers.
The next four terms are the compression and bending energies for an
anisotropic 3D smectic with layers parallel to the $xz$ plane.  The
bending energy is anisotropic due to the presence of the DNA columns.
The final term is a coupling of the nonlinear strains $u_{yy}$ and
$u_{zz}$.

The form of the nonlinear strains depends on whether Eulerian or 
Lagrangian coordinates are used\cite{Lubensky}.  We find it
convenient to use a mixed parameterization in which $x$ and $z$ 
are Eulerian coordinates specifying a position in space and $y=na$ is
a Lagrangian coordinate specifying the layer number.  In
Appendix \ref{app:nonlinstrain}, we derive the nonlinear 
strains $u_{zz}$ and $u_{yy}$ for this mixed parameterization.  To 
quadratic order in gradients of $u_y$ and $u_z$, we find 
\begin{mathletters}
\begin{eqnarray}
\label{uyy}
u_{yy} & = & \partial_y u_y - {1 \over 2}\left[ (\partial_x u_y)^2
+ (\partial_z u_y)^2 - (\partial_y u_y)^2\right] \\
\label{uzz}
u_{zz} & = & \partial_z u_z - {1 \over 2}\Big[ (\partial_x u_z)^2
+ (\partial_z u_z)^2 - (\partial_z u_y)^2\Big]. 
\end{eqnarray}
\end{mathletters}
Note that the nonlinear strain $u_{zz}$ does not contain the shear
strain term proportional to $(\partial_y u_z)^2$.  Thus, layer 
fluctuations do not modify the essential invariance $u_z' 
\rightarrow u_z + f(y)$ of the sliding columnar phase to the 
order considered here\cite{strainfoot}.  In what follows, we will truncate the 
nonlinear strains to
\begin{mathletters}
\begin{eqnarray}
u_{yy} & \approx & \partial_y u_y \\
u_{zz} & \approx & \partial_z u_z -{1 \over 2}(\partial_x u_z)^2
\label{straintrucate}
\end{eqnarray}
\end{mathletters}
since the other nonlinear terms are irrelevant with respect to the sliding
columnar harmonic terms in (\ref{uyuz}).

The goal of this section is to calculate the Grinstein-Pelcovits
renormalization of the eight elastic constants found in the theory of
the sliding columnar phase with lipid bilayer fluctuations.  Since the
nonlinear strains do not introduce a $(\partial_yu_z)^2$ term,
we do not expect the bilayer fluctuations to alter the renormalization
of the SC elastic constants in the simplified theory of the previous
section to lowest order in $B^{yz}$.  We will again use dimensional
regularization to calculate the renormalization.  The format will
closely parallel the previous SC calculation.  We first determine which of the
harmonic terms in (\ref{uyuz}) are relevant and drop irrelevant 
terms.  We then rescale lengths
and fields, ensure that the Hamiltonian retains its unscaled form,
impose boundary boundary conditions on the vertex function, and
calculate the renormalization constants.  The renormalization
constants then determine the scaling form of the vertex function.
 
\subsection{Engineering Dimensions}
\label{sec:layerfluctengineer}

We begin by rescaling the lengths
and the fields in ${\cal H}_b$.  In addition to the rescalings in 
Sec. \ref{sec:scdimanal}, we also rescale $u_y$ according to
\begin{equation}
u_y = L_{u_y} \widetilde{u}_y.
\label{uyscale}
\end{equation}
We first impose the conditions of the previous section, {\it
i.e.}  we set the coefficients of $\widetilde{u}_{zz}^2$ and
$(\partial_{\tilde{x}}
\partial_{\tilde{y}} u_z)^2$ to unity and ensure that both terms 
in the nonlinear strain $u_{zz}$ scale the same way.  As an added
constraint, we set the coefficient of $\widetilde{u}^2_{yy}$ to unity.
These conditions fix
\begin{eqnarray}
L_{u_y} & = & \left({K^z_{xy} \over B^z}\right)^{1/4} {1 \over (B^y)^{1/2}}
\nonumber \\
L_y & = & \left({ (K^z_{xy})^3 \over {B^z}}\right)^{1/4} \nonumber \\ 
L_z & = & L_{u_z}^{-1} = (K^z_{xy} B^z)^{1/4}. 
\label{ly}
\end{eqnarray}

Once we plug in these scaling lengths, the rescaled Hamiltonian becomes 
\begin{eqnarray}
\label{rescaledbilayer}
\widetilde{{\cal H}}_b & = & {1 \over 2} 
\int d^d\widetilde{x} \Big[ \widetilde{u}_{zz}^2 + 
(\partial_{\tilde{x}} \partial_{\tilde{y}} u_z)^2 + w^{-1}
(\partial_{\tilde{x}}^2 \widetilde{u}_z)^2 \\
& + & (\partial_{\tilde{y}} \widetilde{u}_y)^2 + 
2 v(\partial_{\tilde{y}} \widetilde{u}_y) \widetilde{u}_{zz} +
v_1(\partial^2_{\tilde{x}} \widetilde{u}_y)^2 \nonumber \\
& + & v_2 (\partial_{\tilde{x}} \partial_{\tilde{z}} \widetilde{u}_y)^2 +
v_3 (\partial^2_{\tilde{z}} \widetilde{u}_y)^2 \Big] \nonumber
\end{eqnarray}
with 
\begin{eqnarray}
\label{rescaledcoefficients}
w & = & { (B^z)^{1/2} \over K^z_{xx} (K^z_{xy})^{1/2}},~~
v = { B^{yz} \over (B^y B^z)^{1/2}}, \\
v_1 & = & {K^y_{xx}(K^z_{xy})^{3/2}  
\over B^y (B^z)^{1/2} },~~
v_2 = { K^y_{xz} K^z_{xy} \over B^y B^z},~{\rm and} \nonumber \\
v_3 & = & { K^y_{zz} (K^z_{xy})^{1/2} \over (B^z)^{3/2} B^y }. \nonumber
\end{eqnarray}
(It is again necessary to let $x$ represent a $d-2$ displacement with
$d=3-\epsilon$.)  The dimensions of the scaled variables and the $w$
and $v$ coefficients are determined using (\ref{ly}) and the
dimensions of the compression and bending moduli, $[B] = L^{-d}$ and
$[K] = L^{2-d}$.  (Note we have dropped the tildes on the scaled
variables in the following discussion.)  We find
\begin{eqnarray}
\label{fluctscaling}
\left[u_y\right] & = & L^{(1-d)/2},~~\left[v\right] = L^0,~~
\left[v_1\right] = L^{5-d}, \\
\left[v_2\right] & = & L^4,~{\rm and}~
\left[v_3\right] = L^{d+3}, \nonumber 
\end{eqnarray}
while the dimensions of $u_z$, $y$, $z$, and $w$ were
given previously in (\ref{scdimension}).  Note that $v$ does not scale
with length.  Also note that the coefficients $v_1$, $v_2$, and $v_3$
are irrelevant when $d=3$.  We drop the irrelevant terms and arrive at
the following simplified Hamiltonian:
\begin{eqnarray}
{\cal H}_b & = & {1 \over 2} 
\int d^dx \Big[ u_{zz}^2 + 
(\partial_x \partial_y u_z)^2 + w^{-1}
(\partial_x^2 u_z)^2 \nonumber \\
\label{simplifiedtheory}
& + & (\partial_y u_y)^2 +
2 v (\partial_y u_y) u_{zz}\Big].
\end{eqnarray}

\subsection{RG Procedure}
\label{sec:rganalfluct}

The present RG procedure will be similar to those employed 
in sections \ref{sec:smecticrgprocedure} and \ref{sec:scrgprocedure},
except we now have two coupling constants, $w$ and
$v$, instead of one.  We will show that the inclusion of $v$ does not
alter the renormalization of the sliding columnar elastic constants 
to lowest order in $v$.
As before, we rescale the fields and lengths and seek a renormalized
Hamiltonian with the same form as (\ref{simplifiedtheory}).  We
scale $y$, $z$, and $u_z$ as we did previously in
(\ref{scscalerules}) and $u_y$ by $\widetilde{{\cal Z}}^{1/2}$
as follows:
\begin{equation}
u_y({\bf x}) = \widetilde{{\cal Z}}^{1/2} u_y'({\bf x}') = 
\widetilde{{\cal Z}}^{1/2}u_y'(x, {\cal Z}_y y, {\cal Z}^{1/3}z).
\label{Zuyscale}
\end{equation}
The rescaled Hamiltonian ${\cal H}_b'$ 
looks similar to (\ref{hamiltoniansc}) 
with two additional terms due to fluctuations of the bilayers.  We
drop the primes on the variables and find
\begin{eqnarray}
\label{bilayerrescale}
{\cal H}_b & = & {1 \over 2} \int d^dx\Big[ 
{\cal Z} {\cal Z}^{-1}_y u_{zz}^2 
+ {\cal Z}^{1/3} {\cal Z}_y (\partial_x\partial_y u_z)^2 \\
& + & (g\mu^{\epsilon}{\cal Z}_g)^{-1}(\partial^2_xu_z)^2  
+ {\cal Z}^{-1/3}{\cal Z}_y \widetilde{{\cal Z}}(\partial_y u_y)^2 \nonumber \\
& + & 2 \overline{v} 
{\cal Z}_v
(\partial_y u_y) u_{zz} 
\Big], \nonumber 
\end{eqnarray}
where 
\begin{equation}
\overline{v} {\cal Z}_v = v \widetilde{{\cal Z}}^{1/2} {\cal Z}^{1/3}
\label{Zv}
\end{equation}
and ${\cal Z}_g$ was defined previously.

Boundary conditions imposed on the vertex functions $\Gamma_{ij}({\bf
q})$ with $i,j=y,z$ ensure that the Hamiltonian retains its original form
in (\ref{simplifiedtheory}) after rescaling.  The vertex function is
defined by $\Gamma_{ij}({\bf q}) = G^{-1}_{ij}({\bf q})$ with
$G_{ij}({\bf x}) = \langle u_i({\bf x}) u_j(0)\rangle$.  The
conditions imposed on $\Gamma_{zz}$ are identical to those given in
(\ref{rgcondsc1}); these are augmented by two conditions on
$\Gamma_{yz}$ and $\Gamma_{yy}$.
\begin{eqnarray}
\label{bilayerconditions}
\left.{d \Gamma_{yz} \over d(q_y q_z)}\right|_{q_z=\mu^2,q_{x,y}=0} & = & 
2\overline{v} \\
\left.{d \Gamma_{yy} \over dq_y^2}\right|_{q_z=\mu^2,q_{x,y}=0} & = & 1.
\nonumber 
\end{eqnarray}
\begin{figure}
\epsfxsize=1.8truein
\centerline{\epsfbox{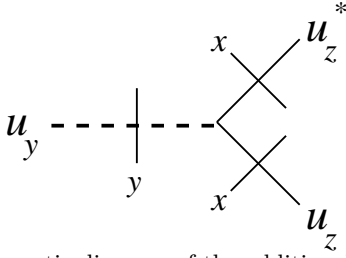}}
\caption{Schematic diagram of the additional relevant nonlinear term
$\partial_y u_y (\partial_x u_z)^2$ generated by the sliding columnar
theory with lipid bilayer fluctuations.  The symbols $x$ and $y$
written adjacent to the dividing lines represent $x$ and $y$
derivatives of the respective fields.  The $u_y$ field is denoted by a
dashed line while $u_z$ is denoted by an unbroken line.}
\label{relevantuyfig}
\end{figure}
Once we impose these conditions on the vertex functions, we  
solve for the ${\cal Z}$'s in terms of the one-loop diagrammatic contributions
$\Sigma_{ij}$, where, for instance, $\Sigma_{zz}$ is the 
one-loop correction to the vertex function $\Gamma_{zz}$. 
The diagrammatic corrections arise from the quadratic term
in $u_{zz}$.  $u_{zz}^2$ generates $\partial_z u_z (\partial_x u_z)^2$,
which was already present in the theory with $u_y=0$.  The coupling of
$u_{yy}$ to $u_{zz}$ generates a new nonlinear term, $\partial_y u_y
(\partial_x u_z)^2$.  This term is shown schematically in
Fig.~\ref{relevantuyfig}.  There are six new one-loop diagrams in
addition to the three diagrams of the rigid sliding columnar theory;
these are shown in Figs.~\ref{Byz} and
\ref{ByzBz}.  The diagrams in Fig.~\ref{Byz} arise from contractions
of $\partial_y u_y (\partial_x u_z)^2$ with itself and the diagrams in
Fig.~\ref{ByzBz} arise from contractions of $\partial_y u_y
(\partial_x u_z)^2$ with $\partial_z u_z (\partial_x u_z)^2$.
\begin{figure}
\epsfxsize=2.2truein
\centerline{\epsfbox{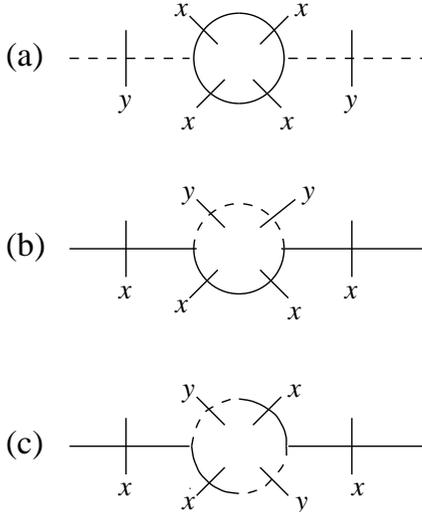}}
\caption{The three diagrams that can be formed by contracting
$\partial_y u_y (\partial_x u_z)^2$ with itself.  The only diagram
that contributes to the renormalization of $B^{y}$ is pictured in (a).
The diagrams in (b) and (c) contribute to the renormalization of both
$K^z_{xx}$ and $K^z_{xy}$.  }
\label{Byz}
\end{figure}
\begin{figure}
\epsfxsize=2.2truein
\centerline{\epsfbox{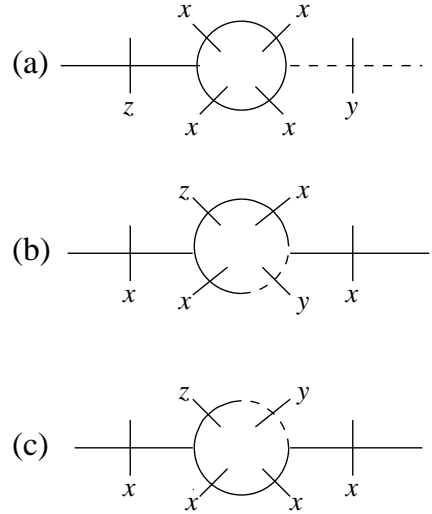}}
\caption{The three diagrams that can be formed by contracting $\partial_z 
u_z (\partial_x u_z)^2$ with $\partial_y u_y (\partial_x u_z)^2$.  The
only diagram that contributes to the renormalization of $B^{yz}$ is
pictured in (a).  The diagrams pictured in (b) and (c)
contribute to the renormalization of both $K^z_{xx}$ and $K^z_{xy}$.}
\label{ByzBz}
\end{figure}
The one-loop diagrammatic corrections $\Sigma_{zz}$ are easy to
calculate since the form of the propagator $G_{zz}$ is unchanged from
its form in the rigid sliding columnar theory.  The form is
not changed, but the compression modulus $B$ is renormalized by a
factor of $1-\overline{v}^2$.  The one-loop diagrammatic corrections
to $\Gamma_{zz}$ are shown below to lowest order in $\epsilon$:
\begin{eqnarray}
\label{zzcorrections}
\left.{d \Sigma_{zz} \over dq_z^2}\right|_{q_z=\mu^2,q_{x,y}=0} & = &
-{g \over 8\pi^2 \epsilon} {1 \over \sqrt{1-\overline{v}^2}}\\
\left.{d \Sigma_{zz} \over d(q_x^2q_y^2)}\right|_{q_z=\mu^2,q_{x,y}=0} & = &
{g \over 24 \pi^2 \epsilon} \sqrt{1-\overline{v}^2} \nonumber \\
\left.{d \Sigma_{zz} \over dq_x^4}\right|_{q_z=\mu^2,q_{x,y}=0} & = &
(g\mu^{\epsilon})^{-1} {g \over 12 \pi^2 \epsilon} \sqrt{1-\overline{v}^2}. 
\nonumber
\end{eqnarray} 
These expressions reduce to those found for the rigid theory when
$\overline{v}=0$.

The calculation of one-loop diagrammatic corrections to $\Gamma_{yz}$
and $\Gamma_{yy}$ is similarly straightforward.
$\Sigma_{yz}$ is given by the diagram in Fig.~\ref{ByzBz}(a).  This
amplitude is proportional to $\overline{v}$ since it is formed by
contracting $\partial_y u_y(\partial_x u_z)^2$ with $\partial_z u_z
(\partial_x u_z)^2$.  $\Sigma_{yy}$ is given by the diagram in
Fig.~\ref{Byz}(a); it must be proportional to $\overline{v}^2$ since
it is formed by contracting $\partial_y u_y (\partial_x u_z)^2$ with
itself.  The one-loop corrections to $\Gamma_{yz}$ and $\Gamma_{yy}$
are given below to lowest order in $\epsilon$:
\begin{eqnarray}
\label{yyyzoneloop}
\left.{d \Sigma_{yz} \over d(q_y q_z)}\right|_{q_z=\mu^2,q_{x,y}=0} & = &
-{g \overline{v} \over 8\pi^2 \epsilon} {1 \over \sqrt{1-\overline{v}^2}}\\
\left.{d \Sigma_{yy} \over dq_y^2}\right|_{q_z=\mu^2,q_{x,y}=0} & = &
-{g \overline{v}^2 \over 8\pi^2 \epsilon} {1 \over \sqrt{1-\overline{v}^2}}. 
\nonumber
\end{eqnarray}

We then use the conditions imposed on the vertex functions 
in (\ref{rgcondsc1}) and (\ref{bilayerconditions}) and the one-loop
diagrammatic corrections in (\ref{zzcorrections}) 
and (\ref{yyyzoneloop}) to find 
the renormalization constants (the ${\cal Z}$'s) in terms of $g$ 
and $\overline{v}$. 
We find that the relations for ${\cal Z}$, ${\cal Z}_y$, and ${\cal Z}_g$
are unchanged to zeroth order in $\overline{v}$.  $\widetilde{{\cal Z}}$ and 
${\cal Z}_v$ also have terms that are independent of $\overline{v}$
as shown below to lowest order in $\epsilon$:
\begin{eqnarray}
\label{ztilde}
\widetilde{{\cal Z}} & \approx & 1 + {g \over 12\pi^2 \epsilon} \\
{\cal Z}_v & \approx & 1 + {g \over 8\pi^2\epsilon}. \nonumber
\end{eqnarray}
The variation of $g$ and $\overline{v}$ with the length scale $\mu$ is
obtained by enforcing that both bare coupling constants do not depend
on $\mu$, {\it i.e.}, we set $\mu dw/d\mu = \mu dv/d\mu = 0$.  
These two requirements determine the recursion
relations for $g$ and $\overline{v}$; we find that $dg/dl$ is
unchanged to lowest order in $\overline{v}$ and
\begin{equation}
{d\overline{v} \over dl} = -{g\overline{v} \over 16\pi^2}.
\label{dvdl}
\end{equation} 
The zeroth order solution for $g$ was found previously in 
(\ref{gsc}); we plug this solution into (\ref{dvdl})
and find 
\begin{equation}
\overline{v}(l) = {\overline{v}_0 \over 
\left[1 + g_0 l/(6\pi^2)\right]^{3/8}},
\label{vofl}
\end{equation}
where $\overline{v}_0 = B^{yz}/\sqrt{B^yB^z}$ and $g_0=\sqrt{B^z/K^z_{xy}}/
K^z_{xx}$.

\subsection{Renormalized Elastic Constants}
\label{sec:bilayerelastic}

We found in the previous two sections that 
the renormalized elastic constants are obtained by solving the 
Callan-Symanzik equation for the renormalized vertex function.  
We find the CS equations for $\Gamma^r_{ij}$ using the 
following scaling equations which relate the bare and renormalized 
vertex functions:
\begin{mathletters}
\begin{eqnarray}
\label{scaling1}
\Gamma_{zz}({\bf q},w,v) & = & {\cal Z}^{-1/3} {\cal Z}_y \Gamma^r_{zz}
({\bf q}',g,\overline{v},\mu) \\  
\label{scaling2}
\Gamma_{yy}({\bf q},w,v) & = & \widetilde{{\cal Z}}^{-1} {\cal Z}_y^{-1}
{\cal Z}^{1/3} \Gamma^r_{yy}
({\bf q}',g,\overline{v},\mu)\\  
\label{scaling3}
\Gamma_{yz}({\bf q},w,v) & = & \widetilde{{\cal Z}}^{-1/2} {\cal Z}_y 
\Gamma^r_{yz} ({\bf q}',g,\overline{v},\mu).   
\end{eqnarray}
\end{mathletters}
Eq. (\ref{scaling1}) yields a CS equation identical to (\ref{CSsc}) to
lowest order in $\overline{v}$, and thus the renormalized elastic
constants $B^z({\bf q})$, $K^z_{xx}({\bf q})$, and $K^z_{xy}({\bf q})$
are identical to those obtained in (\ref{scanswer}) using the $u_y=0$
theory.  The fact that the elastic constants are identical to zeroth
order in $\overline{v}$ is a consequence of the fact that the
nonlinear term proportional to $\overline{v}$ does not introduce any
harmonic terms that were not already present in the theory without $u_y$
fluctuations.  We also find that the coefficient of
$\Gamma^r_{yy}({\bf q}')$ is unity to lowest order $\overline{v}$, and
hence the vertex function $\Gamma_{yy}$ does not rescale.  
As a result, $B^{y} = B^{y}(l=0)$ plus higher order terms in
$\overline{v}$.

We do, however, find a nontrivial renormalization
of $B^{yz}$.  The scaling relation in (\ref{scaling3}) leads to a CS
equation for $\Gamma^r_{yz}$ with a similar form to the one found in
(\ref{CSsc}).  We find
\begin{eqnarray}
\label{CSbilayer}
\Big[ \mu {\partial \over \partial \mu} -{\widetilde{\eta}(g) \over 2}
- {\eta(g) \over 3}\left(q_z {\partial \over \partial q_z}\right) 
\\ + \eta_y(g)
\left(1 - q_y {\partial \over \partial q_y} \right) + 
\beta(g) {\partial
\over \partial g} \Big]\Gamma_r = 0 \nonumber
\end{eqnarray}
to zeroth order in $\overline{v}$, where
\begin{equation} 
\widetilde{\eta}(g) = \beta(g){d(\ln\widetilde{{\cal Z}}) \over dg} =
{g \over 12 \pi^2}
\label{etatilde}
\end{equation}
and $\eta$ and $\eta_y$ were defined previously.  The solution to 
(\ref{CSbilayer})
can be transcribed from (\ref{scsolution}) and is displayed below:
\begin{eqnarray}
\label{bilayerscsol}
& & \Gamma^r_{yz}({\bf q},g,\overline{v}(g),\mu) = 
\exp\left[
\int^l_0\left({\widetilde{\eta} \over 2} -\eta_y\right)dl'\right] \times \\
& & \Gamma^r_{yz}\left(q_x, \exp\left[\int_0^l \eta_y dl'\right] q_y, 
\exp\left[{1 \over 3} \int_0^l \eta dl'\right] q_z, g, \mu_0\right). \nonumber
\end{eqnarray}
Since $\eta$, $\eta_y$, and $\widetilde{\eta}$ scale as $1/l$, the integrals
in the arguments of the exponentials scale logarithmically with $l$.
Thus, the exponentials yield power-laws in $g$, and we find, for example,
\begin{equation}
\exp\left[
\int^l_0\left({\widetilde{\eta} \over 2} -\eta_y\right)dl'\right] = 
\left[{g(l) \over g_0}\right]^{5/8}.
\end{equation}
The renormalized vertex function in (\ref{bilayerscsol}) obeys a
scaling form analogous to the one obeyed by the renormalized
sliding columnar vertex function in (\ref{scalingsolution}).  We find
\begin{eqnarray}
\label{bilayerscalingsolution}
& & \Gamma^r_{yz}({\bf q},g,\mu) = 
b^{-3} \left[g/g_0\right]^{5/8} \times \\
& & \Gamma^r_{yz}\left(bq_x, bq_y 
\left[g/g_0\right]^{-3/8}, b^2q_z \left[g/g_0\right]^{1/8},
g, \mu_0b\right), \nonumber
\end{eqnarray}
where the $b^{-3}$ prefactor is present because $y$ scales as $b$ and
$z$ scales as $b^2$.  We then choose $b = \mu_0^{-1} = [q_z^2 + q_x^2q_y^2
+ w^{-1} q_x^4]^{-1/4}
\equiv [h({\bf q})]^{-1}$ to match the conventions of the previous section,
substitute (\ref{gsc}) for $g/g_0$, and return to variables with
dimension.  The renormalized vertex function becomes
\begin{equation}
\label{bilayeranswer}
\Gamma^r_{yz}({\bf q}) = 2B^{yz}\left[1 + {g_0 \over 6\pi^2}\ln\left[
{\overline{\mu} \over \overline{h}({\bf q})}\right]\right]^{-3/4} q_yq_z,
\end{equation}
where $\overline{\mu}$ and $\overline{h}({\bf q})$ were defined 
previously.  The renormalized elastic constant $B^{yz}({\bf q})$ is 
the coefficient of $q_yq_z$ in the above expression.  Therefore, we
find that both $B^z$ and $B^{yz}$ scale to zero logarithmically 
with ${\bf q}$ at long wavelengths defined by $\overline{h}({\bf q})
\ll \Lambda^{1/2} \exp[-6\pi^2/g_0]$.

\section{Conclusion}
\label{sec:conclusion}

We have calculated the Grinstein-Pelcovits renormalization of the
elastic constants for the sliding columnar phase.  We first used a
simplified model of the sliding columnar phase in which the DNA
columns were prevented from fluctuating perpendicular to the lipid
layers.  We found that the elastic constants scaled as powers of
$\ln[1/q]$ at long wavelengths.  In particular, we found that the
compression modulus $B$ scales to zero and the rotation and bending
moduli $K_y$ and $K$ scale to infinity as $q$ tends to zero.  We then
added in perpendicular fluctuations of the columns perturbatively and
found that the above results were unchanged to lowest order in the
coupling between strains parallel and perpendicular to the lipid
layers.  We employed dimensional regularization in our RG analysis of
the sliding columnar phase to ensure rotational invariance.  RG
schemes that break rotational invariance, such as the momentum-shell
technique, did not yield correct results.
 
\acknowledgments

We thank R.~D. Kamien for helpful comments.
This work was supported in part by the National Science Foundation
under grant DMR97--30405.

\appendix

\section{Evaluation of the 3D Smectic One-Loop Diagrams}
\label{app:smecticloop}

Our task in this Appendix is to calculate $\Sigma({\bf q})$ defined in
Sec. \ref{sec:smecticrgprocedure} as the one-loop diagrammatic
corrections to $\Gamma({\bf q})$, the vertex function for the 3D
smectic.  These corrections arise from the nonlinear terms in the
Hamiltonian in (\ref{scaled}).  The two nonlinear terms are
$\partial_zu_z({\mbox{\boldmath{$\nabla$}}}_{\perp} u)^2/2$ and
$({\mbox{\boldmath{$\nabla$}}}_{\perp} u)^4/8$ (shown schematically in
Fig.~\ref{relevantfig}), and only contractions of the former
contribute to the renormalization to one-loop order. The three
possible contractions are shown in Fig.~\ref{smecticfig}.  The 
diagrammatic corrections $\Sigma({\bf q})$ can be expressed as 
\begin{equation} 
\Sigma({\bf q}) = \Pi_1({\bf q}) q_z^2 + \Pi_2({\bf q}) q_{\perp}^4 
\equiv \Sigma_1({\bf q}) + \Sigma_2({\bf q}).
\label{smselfenergy}
\end{equation}
Note that we have separated the $q_z^2$ and $q_{\perp}^4$ dependence 
of $\Sigma({\bf q})$ so that to lowest order in ${\bf q}$ 
\begin{equation}
\left.{d \Sigma \over dq_z^2}\right|_{q_z = \mu^2, q_{\perp}=0} =
\left.{d \Sigma_1 \over dq_z^2}\right|_{q_z = \mu^2, q_{\perp}=0}
\label{sep1}
\end{equation}
and
\begin{equation}
\left.{d \Sigma \over dq_{\perp}^4}\right|_{q_z = \mu^2, q_{\perp}=0} =
\left.{d \Sigma_2 \over dq_{\perp}^4}\right|_{q_z = \mu^2, q_{\perp}=0}.
\label{sep2}
\end{equation}
The contributions of $d\Sigma_2/dq_z^2$ to $d\Sigma/dq_z^2$ and 
of $d\Sigma_1/dq_{\perp}^4$ to $d\Sigma/dq_{\perp}^4$ at the 
special point $q_z=\mu^2$ and $q_{\perp}=0$ are higher order in 
$\epsilon$ than the contributions in (\ref{sep1}) and (\ref{sep2}).
We begin by calculating $\Sigma_1({\bf q})$.

\begin{figure}
\epsfxsize=2.2truein
\centerline{\epsfbox{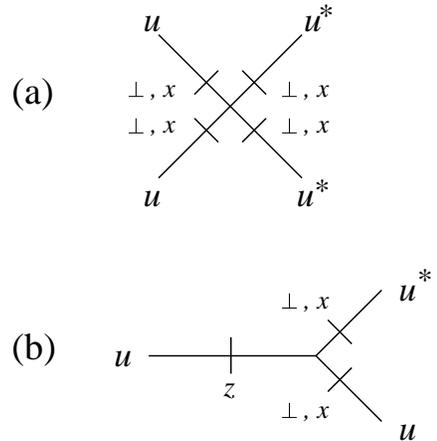}}
\caption{Schematic representation of the two relevant nonlinear terms in 
both the 3D smectic and sliding columnar elasticity
theories.  The perpendicular derivatives (${\perp}$) correspond to the
3D smectic theory and the $x$ derivatives to the sliding columnar
theory.  The term $(\partial_{\perp,x} u)^4$ is pictured in (a) and
the term $(\partial_z u) (\partial_{\perp,x} u)^2$ is pictured in (b).
The symbols ${\perp}$, $x$, and $z$ represent ${\perp}$, $x$, and $z$
derivatives of the $u$ field.  The diagram with four $u$
fields in (a) does not contribute to the renormalization to one-loop
order; only contractions of (b) with itself contribute.}
\label{relevantfig}
\end{figure}

\begin{figure}
\epsfxsize=2.4truein
\centerline{\epsfbox{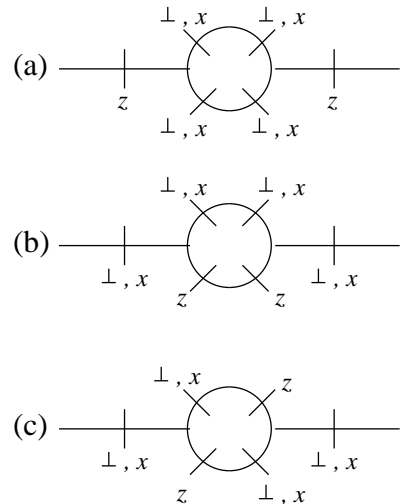}}
\caption{The three one-loop diagrams that contribute to the 
renormalization of the 3D smectic and sliding columnar
elastic constants.  These diagrams are formed by contracting
$\partial_z u (\partial_{\perp, x} u)^2$ with itself.  The diagram in (a)
contributes terms proportional to $q_z^2$ since a factor of $q_z$ is
on each external leg.  The diagrams in (b) and (c) contribute terms
proportional to $q_{\perp}^4$ in the 3D smectic theory and terms
proportional to $q_x^2q_y^2$ and $q_x^4$ in the sliding columnar
theory since these diagrams have $q_{\perp}^2$ or $q_x^2$ 
on the external legs.}
\label{smecticfig}
\end{figure}

\subsection{Calculation of $\Sigma_1({\bf q})$}
\label{app:smecticb}

The diagram in Fig.~\ref{smecticfig}(a) alone contributes to
$\Sigma_1({\bf q})$ since it is the only one with $q_z^2$ on the
external legs.  To evaluate the integrals in the perturbation theory,
we use dimensional regularization, {\it i.e} we take $d = 3 -
\epsilon$, set the cutoff to infinity, and look for the $1/\epsilon$
terms.  $\Sigma_1({\bf q})$ is obtained by calculating the $q_z^2$
contribution from the following integral:
\begin{eqnarray}
\Sigma_1({\bf q}) & = & -{q_z^2 \over 2}\int_{-\infty}^{\infty}
{d^{3 - \epsilon}k \over (2\pi)^{3-\epsilon}} \Big[ (q_{\perp} + k_{\perp})_i
(q_{\perp} + k_{\perp})_j \times \nonumber \\
\label{bsmamplitude}
& & {k_{\perp}}_i {k_{\perp}}_j G({\bf k} + {\bf q})G(-{\bf k}) \Big],
\end{eqnarray}
where $i,j = x,y$ and
\begin{equation}
G({\bf q}) = {1 \over 
q_z^2 + w^{-1} q_{\perp}^4}.
\label{harmpropsm}
\end{equation}
The coefficient of the $q_z^2$ term in (\ref{bsmamplitude}) is $\Pi_1({\bf
q})$.  We can then approximate $\Sigma_1({\bf q})$ by writing
$\Sigma_1({\bf q}) = q_z^2 \Pi_1(q_{\perp}=0, q_z)$
plus higher order terms in $q_{\perp}$ that vanish when we apply the
boundary condition in Eq. (\ref{condition1}).  We obtain $\Pi_1(q_z)$
by setting $q_{\perp}=0$ in the integral on the right hand side of
(\ref{bsmamplitude}).

To evaluate the integral, we first combine the denominators of
$G({\bf k} + {\bf q})$ and $G(-{\bf k})$ employing the following
identity:
\begin{eqnarray}
& & {1 \over (k_z + q_z)^2 + w^{-1}k_{\perp}^4} \times {1 \over
\Big[k_z^2 + w^{-1}k_{\perp}^4\Big]} = \nonumber \\
& &\int^1_0 dx {1 \over 
\Big[ (k_z + x q_z)^2 + x(1 - x)q_z^2
+ w^{-1}k_{\perp}^4\Big]^2}.
\label{smfeynman}
\end{eqnarray}
We then change variables to $k_z' = k_z + xq_z$ and perform the
integration over $k_z'$.  We find that $\Sigma_1({\bf q})$ can be
written in terms of the integral $J(4,3,x,q_z)$ with $J(s,v,x,q_z)$
defined by
\begin{eqnarray}
J(s,v,x,q_z) & = & \int_0^{\infty} dk_{\perp} k_{\perp}^{1 - \epsilon}
{k_{\perp}^s
\over \left[ x(1-x)q_z^2 + w^{-1} k_{\perp}^4\right]^{v/2}} 
\nonumber \\
& = & {w^{v/2} \over 4\Gamma(v/2)} \Gamma
\left({1 \over 4}(2v - s - 2 + \epsilon)\right) \nonumber \\
& & \times \Gamma\left({1 \over 4}(
s + 2 - \epsilon)\right) \nonumber \\
& & \times \left[x(1-x)w q_z^2
\right]^{(s - 2v + 2 - \epsilon)/4},
\label{smintegral}
\end{eqnarray}
where $\Gamma(x)$ is the gamma function evaluated at $x$.  The
expression for $\Sigma_1({\bf q})$ is simple when expressed in terms
of the integral $J(4,3,x,q_z)$; we find 
\begin{equation}
\Sigma_1({\bf q}) = -{q_z^2 \over 16 \pi} \int_0^1 dx~J(4,3,x,q_z).
\label{smfinalstep}
\end{equation}
From (\ref{smintegral}) we know that the most dominant term in
$J(4,3,x,q_z)$ scales as $1/\epsilon$ and thus
\begin{equation}
\Sigma_1({\bf q}) = -{w^{3/2} \over 16 \pi \epsilon} q_z^2 
\left(w q_z^2\right)^{-\epsilon/4} 
\label{smb}
\end{equation}
plus higher order terms in $\epsilon$.  We can also write $\Sigma_1({\bf q})$
as  
\begin{equation}
\left.{d \Sigma_1({\bf q}) \over dq_z^2}\right|_{q_z = \mu^2, q_{\perp}=0}
= -{g \over 16 \pi \epsilon}
\label{qzvertex}
\end{equation}
when we replace $w$ by $(g\mu^{\epsilon})^{2/3}$.

\subsection{Calculation of $\Sigma_2({\bf q})$}
\label{app:smectick}

$\Sigma_2({\bf q})$ is determined by calculating the
$q_{\perp}^4$ contributions from the diagrams in Figs.~\ref{smecticfig}
(b) and (c).  $\Sigma_2({\bf q})$ is the $q_{\perp}^4$ part of the 
the following integral:  
\begin{eqnarray}
\Sigma_2({\bf q}) & = &  
-{q_{\perp}}_i{q_{\perp}}_j \int {d^{3 - \epsilon}k \over (2 \pi)^{3-\epsilon}}
~\Big[ (k_z + q_z)^2 {k_{\perp}}_i {k_{\perp}}_j + \nonumber \\
& & (k_z + q_z) (k_{\perp} + q_{\perp})_j {k_{\perp}}_i k_z \Big] 
\times \nonumber \\
& & G({\bf k} + {\bf q})G(-{\bf k}).
\label{Ksm}
\end{eqnarray}
The $q_{\perp}^4$ contributions come from expanding $G({\bf k+q})$ to
second order in $q_{\perp}$; we see from (\ref{Ksm}) that we need both
the first and second order terms in the expansion.  The coefficient of
the $q_{\perp}^4$ term in the above expansion is $\Pi_2(q_{\perp}=0,
q_z)$, and thus $\Sigma_2({\bf q}) = q_{\perp}^4 \Pi_2(q_z)$ plus
higher order terms in $q_{\perp}$ that vanish when we apply the 
boundary condition in (\ref{condition2}).

The first and second terms in the integrand of (\ref{Ksm}) correspond
to the diagrams in Figs.~\ref{smecticfig} (b) and (c), respectively.
We break up the integral so that $\Sigma_2({\bf q}) =
\Sigma^b_2({\bf q}) + \Sigma^c_2({\bf q})$, and we first calculate
$\Sigma^b_2({\bf q})$.
\begin{eqnarray}
\Sigma^b_2({\bf q}) & = & -{1 \over 2} {q_{\perp}}_i {q_{\perp}}_j 
q_{{\perp}_l} q_{{\perp}_m} \times \\ 
& & \int^{\infty}_{-\infty} {dk_z \over 2\pi}
\int {d\Omega \over (2\pi)^{2-\epsilon}} 
d k_{\perp} k^{1 -\epsilon}_{\perp} \nonumber \\ 
\label{1kb}
& & \left[(k_z + q_z)^2 {k_{\perp}}_i {k_{\perp}}_j G(-{\bf k})
\left.{d^2G({\bf k} + {\bf q}) \over dq_{{\perp}_l} 
dq_{{\perp}_m}}\right|_{q_{\perp}=0}\right], \nonumber
\end{eqnarray}
where $\Omega$ is the solid angle in $2 -\epsilon$ dimensions and the
second derivative of $G$ gives the coefficient of the quadratic term
in the expansion of $G({\bf k} + {\bf q})$.  We then remove the
angular dependence by integrating over $\Omega$ and using the
following two identities:
\begin{equation}
\int {d\Omega \over (2\pi)^{2-\epsilon}}~{k_{\perp}}_i{k_{\perp}}_j 
= {S_{2-\epsilon} \over 2-\epsilon} k_{\perp}^2\delta_{ij}
\label{twoindexes}
\end{equation}
and
\begin{eqnarray}
\label{fourindexes}
& & \int {d\Omega \over (2\pi)^{2-\epsilon}}~{k_{\perp}}_i{k_{\perp}}_j
{k_{\perp}}_l{k_{\perp}}_m = \\
& & {S_{2-\epsilon} \over 
(2-\epsilon)^3} k_{\perp}^4\big(\delta_{ij}\delta_{lm} +
\delta_{il}\delta_{jm} + \delta_{im}\delta_{jl}\big), \nonumber
\end{eqnarray}
where $\delta_{ij}$ is the Kronecker delta and $S_{d} =
\Omega/(2\pi)^d = 2\pi^{d/2}/((2\pi)^d\Gamma(d/2))$ with $d=2-\epsilon$.  
We are interested in the lowest order terms in $\epsilon$ and hence
will use $S_{2-\epsilon} \approx (2\pi)^{-1}$ below.  We then change
variables to $k_z' = k_z + q_z$ and combine the denominators of
$G(-{\bf k})$ and $G({\bf k} + {\bf q})$ using an identity similar to
(\ref{smfeynman}).
\begin{eqnarray}
& & {1 \over (k_z - q_z)^2 + w^{-1}k_{\perp}^4} \times {1 \over
\Big[k_z^2 + w^{-1}k_{\perp}^4\Big]^n} = \\
\label{f2kysm}
& &\Gamma(n+1) \int^1_0 dx {f_n(x)\over 
\Big[ (k_z - x q_z)^2 + x(1 - x)q_z^2
+ w^{-1}k_{\perp}^4\Big]^{n+1}}, \nonumber 
\end{eqnarray}
where $n=2,3$ and 
\begin{equation}
f_n(x) = \left\{
	\begin{array}{ll}
		1 - x, & n = 2\\
		(1 - x)^2/2, & n=3. \\
	\end{array}
	\right.
	\label{fsmrelation}
\end{equation}
We change variables again to $k_z'' = k_z + xq_z$ and 
integrate over $k_z''$; we find that $\Sigma^b_2({\bf q})$ can
be written in terms of the integrals $J(s,v,x,q_z)$ defined 
previously in (\ref{smintegral}):
\begin{eqnarray}
\Sigma^b_2({\bf q}) = -{w^{-1} 
\over 32\pi} q_{\perp}^4 \int_0^1 dx~
\Big[ -5(1-x)J(4,3,x,q_z)  \nonumber \\
-15 x^2(1-x)q_z^2J(4,5,x,q_z) \nonumber \\
+ 9w^{-1}(1-x)^2J(8,5,x,q_z) \nonumber \\
+ 45w^{-1}x^2(1-x)^2q_z^2J(8,7,x,q_z)\Big].
\label{kbsmamplitude}
\end{eqnarray}
$J(4,3,x,q_z)$ and $J(8,5,x,q_z)$ have terms proportional to
$1/\epsilon$ but $J(4,5,x,q_z)$ and $J(8,7,x,q_z)$ do not.  We keep the terms
that are proportional to $1/\epsilon$ and drop the others.  In the
last step we perform the $x$ integration and find
\begin{equation}
\Sigma^b_2({\bf q}) = -{w^{1/2} \over 64\pi \epsilon} 
q_{\perp}^4 (w q_z^2)^{-\epsilon/4}
\label{kbsmfinal}
\end{equation}
plus higher order terms in $\epsilon$.    

We next obtain $\Sigma^c_2({\bf q})$ by calculating the 
$q_{\perp}^4$ contributions from the diagram in Fig.~\ref{smecticfig}(c).
$\Sigma^c_2({\bf q})$ can be written in terms of the following integral:
\begin{eqnarray}
\Sigma^c_2({\bf q}) =  
-{q_{\perp}}_i {q_{\perp}}_j  
\int {d\Omega \over (2\pi)^{2-\epsilon}} dk_{\perp} k^{1-\epsilon}_{\perp}
\int_{-\infty}^{\infty} {dk_z \over 2\pi} \nonumber \\
\Big[k_z(k_z + q_z)
G(-{\bf k})\Big[{k_{\perp}}_i{q_{\perp}}_jq_{{\perp}_l} 
\left.{dG({\bf k} + {\bf q})
\over dq_{{\perp}_l}}\right|_{q_{\perp} = 0} \nonumber \\
+ {k_{\perp}}_i{k_{\perp}}_j {q_{{\perp}_l} q_{{\perp}_m} \over 2}
\left.{d^2G({\bf k} + {\bf q}) 
\over dq_{{\perp}_l} dq_{{\perp}_m}}\right|_{q_{\perp}=0} \Big]\Big].  
\label{kcsm}
\end{eqnarray}
The first and second derivatives of $G$ give the coefficients of the
linear and quadratic terms in $q_{\perp}$ in the expansion of $G({\bf k} +
{\bf q})$.  We then follow a procedure similar to the one employed to
find $\Sigma^b_2({\bf q})$, {\it i.e.}, we change variables to $k_z'
= k_z + q_z$, combine the denominators of $G({\bf k} + {\bf q})$ and
$G(-{\bf k})$, and integrate over $\Omega$.  The remaining integrals
in (\ref{kcsm}) are over $k_{\perp}$ and $x$.  We then integrate over
$k_{\perp}$ and write $\Sigma^c_2({\bf q})$ in terms of $J(s,v,x,q_z)$; 
we find
\begin{eqnarray}
\Sigma^c_2({\bf q}) & = &
-{w^{-1} \over 32\pi} q_{\perp}^4
\int_0^1
dx~\Big[ -9(1-x)J(4,3,x,q_z) \nonumber \\
& & +27x(1-x)^2q_z^2J(4,5,x,q_z) 
\nonumber \\
& & + 9w^{-1}(1-x)^2J(8,5,x,q_z) \nonumber \\
& & -45w^{-1}x(1-x)^2 q_z^2J(8,7,x,q_z)\Big].
\label{kcinter}
\end{eqnarray}
Only $J(4,3,x,q_z)$ and $J(8,5,x,q_z)$ have terms proportional to
$1/\epsilon$.  We keep these terms and perform the integration over
$x$ to find
\begin{equation}
\Sigma^c_2({\bf q}) = {3 w^{1/2} \over 64 \pi \epsilon} 
q_{\perp}^4 (w q_z^2)^{-\epsilon/4}.
\label{kcsmfinal}
\end{equation}
We obtain $\Sigma_2({\bf q})$ by adding $\Sigma^b_2({\bf q})$ and
$\Sigma^c_2({\bf q})$ in (\ref{kbsmfinal}) and (\ref{kcsmfinal}) to yield
\begin{equation}
\left.{d\Sigma_2({\bf q}) \over dq_{\perp}^4}\right|_{q_z = \mu^2,
q_{\perp}=0} = (g \mu^{\epsilon})^{-2/3} {g \over 32\pi \epsilon},
\label{ksmfinal}
\end{equation}
once we set $w = (g\mu^{\epsilon})^{2/3}$ and ignore higher order
terms in $\epsilon$.

\section{Evaluation of the Sliding Columnar Loop Diagrams}
\label{app:slideloop}

The aim of this Appendix is to calculate $\Sigma({\bf q})$, the
one-loop diagrammatic corrections to the vertex function for the
sliding columnar phase.  The rotationally invariant theory given in
(\ref{scaledsc}) contains two relevant nonlinear terms, $\partial_z
u_z (\partial_x u_z)^2$ and $(\partial_x u_z)^4$.  These terms are
pictured schematically in Fig.~\ref{relevantfig}.  From this figure we
see that only contractions of $\partial_z u_z (\partial_x u_z)^2$
renormalize the elastic constants to one-loop order.  The three
possible contractions are shown in Fig.~\ref{smecticfig}.
$\Sigma({\bf q})$ has $q_z^2$, $q_x^2 q_y^2$, and $q_x^4$
contributions, and we will calculate each separately below.
To do this, we express $\Sigma({\bf q})$ as
\begin{eqnarray}
\label{scdefine}
\Sigma({\bf q}) & = & \Pi_1({\bf q}) q_z^2 + \Pi_2({\bf q}) q_x^2 q_y^2
+ \Pi_3({\bf q}) q_x^4 \\
& \equiv & \Sigma_1({\bf q}) + \Sigma_2({\bf q}) + \Sigma_3({\bf q}). \nonumber
\end{eqnarray}
We have separated the $q_z^2$, $q_x^2q_y^2$, and $q_x^4$ dependences
so that, for instance, 
\begin{equation}
\left.{d \Sigma \over dq_x^4}\right|_{q_z = \mu^2, q_{\perp}=0} =
\left.{d \Sigma_3 \over dq_x^4}\right|_{q_z = \mu^2, q_{\perp}=0}.
\label{sep3}
\end{equation}
As in Appendix \ref{app:smecticloop}, we use dimensional regularization to 
calculate the integrals.

\subsection{Calculation of $\Sigma_1({\bf q})$}
\label{app:slideb}

The $q_z^2$ contribution to $\Sigma({\bf q})$ results from squaring the
diagram pictured in Fig.~\ref{relevantfig}(b) and contracting both
pairs of $x$ derivatives.  This leaves $q_z$ on each external leg
as shown in Fig.~\ref{smecticfig}(a).  $\Sigma_1({\bf q})$
is the $q_z^2$ part of the following integral:
\begin{equation}
\Sigma_1({\bf q}) = -{q_z^2 \over 2} \int 
{d^{3 - \epsilon}k \over (2\pi)^{3-\epsilon}} 
\Big[ (q_x + k_x)^2 k_x^2 G({\bf q
+ k})G(-{\bf k}) \Big],
\label{bamplitude}
\end{equation}  
where 
\begin{equation}
G({\bf q}) = {1 \over q_z^2 + q_x^2 q_y^2 + w^{-1}q_x^4}.
\label{scpropagator}
\end{equation}
The coefficient of the $q_z^2$ in the above integral is $\Pi_1({\bf
q})$ and thus $\Sigma_1({\bf q}) = q_z^2 \Pi_1(q_{x,y}=0,q_z)$ plus
higher order terms in $q_x$ and $q_y$ that vanish when we apply
the boundary condition in (\ref{rgcondsc1}).  Thus, $\Sigma_1({\bf q})$
is obtained by setting $q_x=q_y = 0$ in (\ref{bamplitude}).  We find
\begin{eqnarray}
\Sigma_1({\bf q}) & = & -{w^{-1/2} \over 2}
{q_z^2 \over (2 \pi)^{3-\epsilon}} \\
\label{step1}
& & \times \int dk_x dk_z d^{1-\epsilon}k_y~
\Big[{k_x^4 \over k_z^2 + w^{-1} k_x^2 k_{\perp}^2} \nonumber \\ & &
\times {1 \over (q_z + k_z)^2 + w^{-1} k_x^2 k_{\perp}^2}\Big], \nonumber
\end{eqnarray}
where we have changed variables to $k_y = w^{-1/2}k_y'$ and dropped
the prime.  The first step in evaluating this integral is to combine
the two denominators in (\ref{step1}) using the identity in
(\ref{smfeynman}) with $k_{\perp}^4$ replaced by $k_x^2k_{\perp}^2$.
We then perform the integration over $k_z$ and find that
$\Sigma_1({\bf q})$ can be written in terms of the integral $I(4,0,3,x,q_z)$,
where
\begin{eqnarray}
I(s,t,v,x,q_z)  & = &
\int^{\infty}_0 dk_x dk_y~{ k_x^s k_y^{t - \epsilon} \over 
\Big[x(1 - x) q_z^2 + 
w^{-1}k_x^2 k_{\perp}^2\Big]^{v/2}} \nonumber \\
& = & {w^{v/2} \over 8 \Gamma(v/2)} \Gamma\left({1 \over 2}(t + 1 - 
\epsilon)\right) 
\Gamma\left({1 \over 4}(s - t + \epsilon)\right)  \nonumber \\
& & \times \Gamma\left({1\over 4}(2v - t - s - 2 + \epsilon)
\right) \nonumber \\
&  & \times \Big[x(1 - x) w q_z^2 \Big]^{
(s + t - 2v + 2 - \epsilon)/4}.
\label{general}
\end{eqnarray}
We give the most general form for the integrals over $k_x$ and $k_y$
since we will need these integrals later when we calculate $\Sigma_2({\bf q})$
and $\Sigma_3({\bf q})$.  We find 
\begin{equation}
\Sigma_1({\bf q}) = {-w^{-1/2} \over 8\pi^2} q_z^2 \int_0^1
dx~I(4,0,3,x,q_z).
\label{interstep}
\end{equation}
and 
\begin{equation}
\Sigma_1({\bf q}) = - {w \over 8 \pi^2 \epsilon} q_z^2 (w q_z^2)^{-\epsilon/4}
\label{appfinalstep}
\end{equation}
since $I(4,0,3,x,q_z) \propto 1/\epsilon$.  We then set $w = g\mu^{\epsilon}$
to find $\Sigma_1({\bf q})$ as a function of $g$,
\begin{equation}
\left.{d\Sigma_1({\bf q}) \over dq_z^2}\right|_{q_z=\mu^2, q_{x,y}=0} 
= - {g \over 8 \pi^2 \epsilon}. 
\label{gammabg}
\end{equation}

\subsection{Calculation of $\Sigma_2({\bf q})$}
\label{app:slideky}

Both the $q_x^2q_y^2$ and $q_x^4$ contributions to $\Sigma({\bf q})$
come from the diagrams with $x$ derivatives on the external legs.  The
two contributing diagrams are shown in Figs.~\ref{smecticfig} (b) and
(c).  Their sum is given by 
\begin{eqnarray}
\label{KK}
& & {\rm Sum} = \\
& & -q_x^2
\int {d^{3 - \epsilon}k \over (2\pi)^3} \Big[ (k_z + q_z)^2 k_x^2 + 
(q_z + k_z)(q_x + k_x)k_zk_x\Big]  \nonumber \\
& & \times G({\bf k} + {\bf q})G(-{\bf k}).  \nonumber
\end{eqnarray}
We find the $q_x^2q_y^2$ terms by expanding $G({\bf k} + {\bf q})$ to
second order in $q_y$.  We see that only the quadratic term in the
expansion contributes.  Higher order terms will vanish when we apply
the second boundary condition in (\ref{rgcondsc1}).  We then follow a
procedure similar to the one employed to calculate the $q_{\perp}^4$
contribution to the $3$D smectic vertex function in Appendix
\ref{app:smecticloop}.  We find that $\Sigma_2({\bf q})$ can be
written in terms of the integrals $I(s,t,v,x,q_z)$ as shown below:
\begin{eqnarray}
\Sigma_2({\bf q}) & = & -{w^{-1/2} \over 8 \pi^2} q_x^2q_y^2
\int^1_0 dx
\Big[ -2(1 - x)I(4,0,3,x,q_z) + \nonumber \\ 
& & 6w^{-1}(1 -x)^2 I(6,2,5,x,q_z) - \nonumber \\
& & 3xq_z^2(2x-1)(1 -x)I(4,0,5,x,q_z) + \nonumber \\
& & 15w^{-1}xq_z^2(2x-1)(1-x)^2 I(6,2,7,x,q_z)\Big].
\label{endstep}
\end{eqnarray}
We look for the leading order terms in $\epsilon$ in
(\ref{endstep}); $I(4,0,3,x,q_z)$ and $I(6,2,5,x,q_z)$ have leading order
terms proportional to $1/\epsilon$ while $I(4,0,5,x,q_z)$ and $I(6,2,7,x,q_z)$
do not and are dropped.  After integrating (\ref{endstep}) over $x$ we
obtain
\begin{equation}
\Sigma_2({\bf q}) = {w \over 24 \pi^2 \epsilon} 
q_x^2q_y^2 (wq_z^2)^{-\epsilon/4}
\label{attkyamplitude}
\end{equation}
and 
\begin{equation}
\left.{d \Sigma_2 \over d(q_x^2q_y^2)}\right|_{q_z = \mu^2, q_{x,y} = 0} 
= {g \over 24 \pi^2 \epsilon}.
\label{gammakyg}
\end{equation}

\subsection{Calculation of $\Sigma_3({\bf q})$}
\label{app:slidek}

$\Sigma_3({\bf q})$ is obtained by calculating the terms proportional
to $q_x^4$ in (\ref{KK}).  We obtain these terms by expanding $G({\bf
k} + {\bf q})$ to second order in $q_x$ and noting that both first and
second order terms in the expansion contribute.  Note that higher
order terms in the expansion will vanish once we apply the third boundary
condition in (\ref{rgcondsc1}).  We calculate the $q_x^4$
contributions from Figs.~\ref{smecticfig} (b) and (c) separately and
define $\Sigma_3({\bf q}) \equiv
\Sigma_3^b({\bf q}) + \Sigma_3^c({\bf q})$.
We first calculate the contribution from Fig.~\ref{smecticfig}(b).
Using the same procedure as the one employed to calculate
the $q_x^2q_y^2$ contribution to $\Sigma({\bf q})$, we find that
$\Sigma_3^b({\bf q})$ can be written in terms of the integral
$I(s,t,v,x,q_z)$.
\begin{eqnarray}
\Sigma^b_3({\bf q}) & = & -{w^{-3/2} \over 8 \pi^2} q_x^4 \int_0^1 dx~
\Big[ -(1-x)(6I(4,0,3,x,q_z) + \nonumber \\
& & I(2,2,3,x,q_z) + 18x^2q_z^2I(4,0,5,x,q_z) +\nonumber \\
& & 3x^2q_z^2I(2,2,5,x,q_z)) + \nonumber \\ 
& & 3w^{-1}(1-x)^2\Big(4I(8,0,5,x,q_z) + \nonumber \\
& & 20x^2q_z^2I(8,0,7,x,q_z) + \nonumber \\ 
& & 4I(6,2,5,x,q_z) + 20x^2q_z^2I(6,2,7,x,q_z) +\nonumber \\
& & I(4,4,5,x,q_z) + 5x^2q_z^2I(4,4,7,x,q_z)\Big) \Big].
\label{Ib}
\end{eqnarray}
We note that three of the integrals in (\ref{Ib}), $I(4,0,5,x,q_z)$,
$I(8,0,7,x,q_z)$, and $I(6,2,7,x,q_z)$, have leading order terms that
scale as $\epsilon^0$ and are dropped.  Two integrals,
$I(2,2,3,x,q_z)$ and $I(4,4,5,x,q_z)$, have $1/\epsilon^2$ as well as
$1/\epsilon$ terms, while the remaining five integrals
$I(4,0,3,x,q_z)$, $I(2,2,5,x,q_z)$, $I(8,0,5,x,q_z)$,
$I(6,2,5,x,q_z)$, and $I(4,4,7,x,q_z)$ have leading order
contributions that scale as $1/\epsilon$.  We collect terms and
perform the $x$ integration to find
\begin{equation}
\Sigma_3^b({\bf q}) = -{1 \over 8 \pi^2 \epsilon} q_x^4 (w q_z^2)^{-\epsilon/4}
\left[
{1 \over \epsilon} + \ln [2] - {1 \over 12}\right].
\label{deltaKb}
\end{equation}
Note that the dominant contribution to $\Sigma_3^b$ is of order
$\epsilon^{-2}$ rather than $\epsilon^{-1}$.  The undesirable
$\epsilon^{-2}$ term and the $\ln[2]/\epsilon$ term will be cancelled
by terms in $\Sigma_3^c$.  The term proportional to $\ln
[2]/\epsilon$ originates from the integrals $I(2,2,3,x,q_z)$ and
$I(4,4,5,x,q_z)$.  This can be seen by expanding $I(4,4,5,x,q_z)$ in
powers of $\epsilon$; we find
\begin{eqnarray}
I(4,4,5,x,q_z) & = & {2 w^{-5/2} \over \epsilon^2} \left(
1 - {\epsilon \over 2} {\Gamma'(5/2) \over \Gamma(5/2)} +
{\epsilon \over 2}{\Gamma'(1) \over \Gamma(1)} \right) \times \nonumber \\
& & \left[x(1-x)wq_z^2\right]^{-\epsilon/4}
\label{log}
\end{eqnarray}
to order ${\cal O}(1/\epsilon)$, where $\Gamma'(x)$ is the derivative
of the gamma function evaluated at $x$.  The logarithm arises from
evaluating the derivative of the gamma function at a half integer.
For example, $\Gamma'(5/2)/\Gamma(5/2) = -\gamma + 8/3 -2\ln[2]$ where
$\gamma$ is the Euler-Mascheroni constant.  

We can also write $\Sigma_3^c({\bf q})$ in terms of the integrals
$I(s,t,w,x,q_z)$.  We obtain
\begin{eqnarray}
\Sigma_3^c({\bf q}) & = & -{w^{-3/2} \over 8 \pi^2} q_x^4
\int_0^1 dx \Big[ (1-x)\Big(-10I(4,0,3,x,q_z) + \nonumber \\
& & 30x(1-x)q_z^2I(4,0,5,x,q_z) - \nonumber \\ 
& & 3I(2,2,3,x,q_z) + 9x(1-x)q_z^2I(2,2,5,x,q_z)\Big) + \nonumber \\
& & 3w^{-1}(1-x)^2\Big( 4I(8,0,5,x,q_z) - \nonumber \\ 
& & 20x(1-x)q_z^2I(8,0,7,x,q_z) + 4I(6,2,5,x,q_z) - \nonumber \\
& & 20x(1-x)q_z^2I(6,2,7,x,q_z) + I(4,4,5,x,q_z) - \nonumber \\
& & 5x(1-x)q_z^2I(4,4,7,x,q_z)\Big)\Big],
\label{kcfinal}
\end{eqnarray}
which becomes
\begin{eqnarray}
\Sigma_3^c({\bf q}) = -{1 \over 8\pi^2 \epsilon} q_x^4 (w q_z^2)^{-\epsilon/4}
\Big[-{1 \over \epsilon} -\ln[2] -{7 \over 12}\Big]
\label{final}
\end{eqnarray}
when only terms proportional to $1/\epsilon^2$ and $1/\epsilon$ are
retained.  We see that when we add (\ref{deltaKb}) to (\ref{final}),
the terms proportional to $1/\epsilon^2$ and $\ln[2]/\epsilon$ cancel
and we are left with
\begin{equation}
\Sigma_3({\bf q}) = {1 \over 12 \pi^2 \epsilon} q_x^4 (w q_z^2)^{-\epsilon/4}
\label{kfinal}
\end{equation}
and
\begin{equation}
\left.{d \Sigma_3({\bf q}) \over dq_x^4}\right|_{q_z = \mu^2,q_{x,y} =0} 
= (g \mu^{\epsilon})^{-1} {g \over 12 \pi^2 \epsilon}.
\label{gammakg}
\end{equation}

\section{Finite Wavenumber Cutoff}
\label{app:cutoff}

In this Appendix we show that employing a finite cutoff leads to
ambiguities when we evaluate the sliding columnar one-loop diagrams.
These diagrams are shown in Fig.~\ref{smecticfig}; (a) contributes to
$\Sigma_1({\bf q})$ and both (b) and (c) contribute to $\Sigma_2({\bf
q})$ and $\Sigma_3({\bf q})$.  The ambiguous result is that we obtain
different answers for $\Sigma({\bf q})$ depending on whether external
momentum $q$ is sent through the top or bottom part of the internal
loop.  The ambiguity develops when momentum $q_x$ appears in the
internal loop and the top and bottom paths through the internal loop
are different.  The diagram that causes this ambiguity is the $q_x^4$
part of Fig.~\ref{smecticfig}(b).  We can see this by calculating the
$q_x^4$ corrections to the vertex function, $\Sigma_3^b({\rm top})$
and $\Sigma_3^b({\rm bot})$, which result from sending 
${\bf k}+{\bf q}$ through the
top(bottom) sections of the internal loop.
\begin{equation}
\Sigma_3^b({\rm top}) = -q_x^2\int_{\Lambda} {d^3 k \over (2\pi)^3}
\left[k_z^2 (k_x + q_x)^2 G(-{\bf k})G({\bf k} + {\bf q})\right],
\label{top}
\end{equation}
and 
\begin{equation}
\Sigma_3^b({\rm bot}) = -q_x^2\int_{\Lambda} {d^3 k \over (2\pi)^3}
\left[k_x^2 (k_z + q_z)^2 G(-{\bf k})G({\bf k} + {\bf q})\right],
\label{bottom}
\end{equation}
where $\Lambda$ is a finite wavenumber cutoff and $G({\bf q})$ was
defined previously in (\ref{scpropagator}).  With $\Lambda \ne
\infty$,
\begin{equation}
\Sigma_3^b({\rm top}) \ne \Sigma_3^b({\rm bot}).
\label{notequal}
\end{equation}
If we employ dimensional regularization instead and send $\Lambda \rightarrow
\infty$, these top and bottom amplitudes are identical.

\section{Derivation of the Nonlinear Strains in the Presence of 
Flexible Membranes}
\label{app:nonlinstrain}

In this appendix, we derive expressions for the nonlinear strains
$u_{yy}({\bf x})$ and $u_{zz}({\bf x})$ introduced in (\ref{uyy}) and
(\ref{uzz}) for the case of flexible membranes.  A complete
description of lamellar DNA-lipid complexes requires separate
coordinates for each membrane and each DNA molecule.  Displacements of
membranes and DNA molecules parallel to the membrane normals
(along the $y$-direction when the membranes are flat) are locked
together.  We can, therefore, model the complexes as a
stack of membranes each with a one-dimensional mass-density wave
representing the DNA lattice just above it.  We employ mixed
Lagrangian-Eulerian variables in which the coordinate $y=na$
specifying the layer or membrane number is a Lagrangian variable and
the coordinates $(x,z) \equiv {\bf r}$ are Eulerian
variables specifying positions in a fixed projection plane.  The
positions of mass points on membrane $n$ are then given by
\begin{equation}
{\bf R}_n({\bf r}) = x {\hat x} + z {\hat z} + 
\left[na + u_y(na,{\bf r})\right]{\hat y}.
\label{memposition}
\end{equation}
The density in membrane $n$ can be expanded as $\rho_n({\bf r}) =
\rho_n^0 + \psi_n({\bf r}) + \psi^*_n({\bf r})$, where 
$\rho_n^0$ is a constant, $\psi_n({\bf r}) = |\psi_n|e^{i\phi_n({\bf r})}$,
and 
\begin{equation}
\phi_n({\bf r}) = k_0\left[z - u_z(na,{\bf r})\right]
\label{phi}
\end{equation}
with $k_0 = 2\pi/d$.  

To construct the strain variable $u_{yy}({\bf x})$ with ${\bf x} =
(y,{\bf r})$, we introduce the distance $l_n({\bf r},{\bf r}')$
between points ${\bf r}$ on membrane $n$ and ${\bf r}'$ on membrane
$n+1$ via
\begin{equation}
l_n^2({\bf r},{\bf r}') = \left| {\bf R}_{n+1}({\bf r}') - {\bf R}_n({\bf r})
\right|^2.
\label{measure}
\end{equation}
The shortest distance between a point ${\bf r}$ on membrane $n$ and any
point on membrane $n+1$ is then 
\begin{equation}
l^2_n({\bf r}) = \min_{{\bf r}'} l^2_n({\bf r},{\bf r}').
\label{mindistance}
\end{equation} 
The strain variable $u_{yy}$ is defined as 
\begin{equation}
u_{yy}({\bf x}) = \lim_{a\rightarrow 0} {1 \over 2a^2}\left(
l^2_{y/a}({\bf r}) - a^2\right).
\label{uystraindef}
\end{equation}
This quantity is by construction invariant with respect to global rotations 
of the entire system.  To evaluate $u_{yy}({\bf x})$, we expand
${\bf R}_{n+1}({\bf r}') - {\bf R}_n({\bf r})$ to lowest order in 
$\delta {\bf r} = {\bf r}' - {\bf r}$ and $a$:
\begin{equation}
{\bf R}_{n+1}({\bf r}') = {\bf R}_n({\bf r}) + a\left(1 + \partial_y
u_y({\bf x})
\right){\hat y} + \delta r^{\mu} {\bf e}_{\mu},
\label{linearorder}
\end{equation}
where $\mu = x,z$, ${\bf e}_{\mu} = \partial_{\mu}{\bf R}_n({\bf x})$ 
is a covariant tangent-plane vector of the $n$th surface,
and $u_y({\bf x}) = u_y(na,{\bf r})$.  Then
\begin{eqnarray}
l^2_n({\bf r},{\bf r}') & = & a^2\left(1+\partial_yu_y\right)^2 +
2a\left(1+\partial_yu_y\right)\delta r^{\mu}\partial_{\mu}u_y \nonumber \\
\label{measureexpand}
& + & g_{\mu \nu} \delta r^{\mu} \delta r^{\nu},
\end{eqnarray}
where $g_{\mu\nu} = {\bf e}_{\mu} \cdot {\bf e}_{\nu}$ is the metric
tensor of the $n$th surface and where we used ${\hat y} \cdot {\bf e}_{\mu} =
\partial_{\mu} u_y$.  We then minimize $l^2_n({\bf r},{\bf r}')$ over
$\delta r^{\mu}$ and obtain
\begin{equation}
\delta r^{\mu} = -a\left(1+\partial_yu_y\right)g^{\mu\nu} \partial_{\nu}u_y
\label{optimalr}
\end{equation}
and
\begin{equation}
l^2_{y/a}({\bf r}) = a^2\left(1 + \partial_yu_y\right)^2\left(1 -
g^{\mu\nu}\partial_{\mu}u_y\partial_{\nu}u_y\right).
\label{intermediate}
\end{equation}
Finally, using $g^{\mu \nu} = (g_{\mu\nu})^{-1}$ where
\begin{equation}
g_{\mu\nu} = \delta_{\mu \nu} + \partial_{\mu} u_y \partial_{\nu} u_y,
\label{metric}
\end{equation}
we obtain
\begin{eqnarray}
\label{finaluyanswer}
u_{yy}({\bf x}) & = & {1\over 2}\left[ {(1 +\partial_yu_y)^2 \over
1 + ({\mbox{\boldmath{$\nabla$}}}u_y)^2} - 1\right] \\
& \approx & \partial_yu_y - {1 \over 2}\left[ (\partial_xu_y)^2 +
(\partial_zu_y)^2 - (\partial_yu_y)^2\right], \nonumber
\end{eqnarray}
with ${\mbox{\boldmath{$\nabla$}}} = (\partial_x,0,\partial_z)$.  It
is straightforward to verify that $u_{yy}({\bf x})=0$ for a uniform
rotation of the entire system.  For example, a rotation of the system
about the $z$ axis by $\theta$ produces strains $\partial_yu_y =
1/\cos\theta - 1$ and $\partial_xu_y = \tan\theta$ which cause
$u_{yy}$ to vanish.

The strain $u_{zz}({\bf x})$ can also be defined in a rotationally invariant
way via
\begin{equation}
u_{zz}({\bf x}) = {1 \over 2k_0^2}\left[k_0^2 - 
g^{\mu\nu}\partial_{\mu}\phi({\bf x})\partial_{\nu}\phi({\bf x})\right],
\label{uzzdef}
\end{equation}
where $\phi({\bf x}) = \phi_{y/a}({\bf r})$ is defined in (\ref{phi}).
To quadratic order in $\partial_{\mu}u_z$ and $\partial_{\mu}u_y$, the
nonlinear strain $u_{zz}$ is
\begin{eqnarray}
u_{zz}({\bf x}) & \approx & \partial_z u_z -{1 \over 2}\Big[
(\partial_xu_z)^2 + (\partial_zu_z)^2 - (\partial_zu_y)^2\Big],
\label{finaluzz}
\end{eqnarray}
where $u_z({\bf x}) = u_z(y,{\bf r})$.


\begin{references}
\bibitem{strainfoot}
The true invariance of the sliding 
columnar phase is that smectic lattices in neighboring galleries can slide
freely relative to each other regardless of the shape of intervening
lipid bilayers.  A complete description of this invariance involves
membrane curvature, which is irrelevant in the RG sense for the 
current calculations.
\end{references}
\end{document}